\documentclass[aps,prb,twocolumn,superscriptaddress]{revtex4-2}

\usepackage[utf8]{inputenc}
\usepackage{mathtools}
\usepackage{amsmath, amsthm, amssymb, amsfonts}
\usepackage{hyperref}
\usepackage{graphicx}
\usepackage{xcolor}

\begin{document}

\title{Combine effect of site dilution and long-range interaction on magnetic and transport properties in the half-filled Hubbard model}

\author{Sudip Mandal}
\affiliation{Theory Division, Saha Institute of Nuclear Physics, A CI of Homi Bhabha National Institute, Kolkata 700064, India}

\author{Sourav Chakraborty}
\affiliation{Theory Division, Saha Institute of Nuclear Physics, A CI of Homi Bhabha National Institute, Kolkata 700064, India}
\affiliation{Ajou Energy Science Research Center, Ajou University, Suwon 16499, Republic of Korea}

\author{Kalpataru Pradhan}
\affiliation{Theory Division, Saha Institute of Nuclear Physics, A CI of Homi Bhabha National Institute, Kolkata 700064, India}

\date{\today}

\begin{abstract}
We investigate the magnetotransport properties of a diluted half-filled one-band
Hubbard model with second-nearest-neighbor hopping on a simple cubic lattice,
aiming to explore the possibility of metallicity in diluted antiferromagnetic
systems. Our semiclassical Monte Carlo calculations reveal an
antiferromagnetic metallic regime in diluted correlated materials. This
unexpected metallic regime naturally leads to a central question: how does
the introduction of dilution into an antiferromagnetic material, especially
with long-range magnetic interactions, induce metallicity---a feature not
commonly associated with antiferromagnets? To address this question, we
demonstrate that when the on-site repulsive Hubbard interaction strength is set to zero
on a percentage of the sites (site dilution), the insulating state weakens
due to percolative conduction among the diluted sites at low temperatures.
Remarkably, this occurs without any significant alteration to the underlying
long-range antiferromagnetic ordering in the system, thereby providing
a pathway to realize antiferromagnetic metals. In addition, we show how the
sublattice-dependent hopping can be exploited to engineer spin-polarized
half-metallic antiferromagnets. Overall, our numerical results collectively
provide a basis for understanding the combined effect of site dilution and
competing interactions, which will assist in the design of new antiferromagnetic
metals for future spintronic applications.
\end{abstract}


\maketitle

\section{Introduction}

In strongly correlated electron systems, antiferromagnetism is traditionally
associated with an insulating state, while ferromagnetism is linked to
metallicity~\cite{Gonzalez, Avella, Liu, Coey, Sachdev}.
The interplay between electron interactions and spin ordering is the basis for
this basic distinction, since antiferromagnetic (AF) materials frequently have a gap
at the Fermi level in the electronic structure that suppresses charge
transport~\cite{Ran, Milloch, Laad}. On the other hand, spin-aligned conduction
channels in ferromagnetic materials tend to favor metallic
behavior~\cite{Brando, Davidson, Lungu}. However, over the past few decades,
quite a few AF metals that contradict the conventional scenario
have been identified~\cite{Siddiqui, Yamashita}. Since they reveal new
perspectives on unusual electronic structures and magnetic properties,
AF metals are especially fascinating for both theoretical and
experimental investigations. More recently, AF metals have been
explored for possible spintronics applications~\cite{Jungwirth, Wang, Baltz, Fukami, DalDin, KhaliliAmiri}.
Combining the advantages of electrical conductivity and zero net magnetism,
particularly with spin polarization at the Fermi level, is a topic of great
scientific interest~\cite{Shim, Guo}. In contrast to ferromagnets, which
generate stray magnetic fields that may disrupt nearby components in the
devices~\cite{Tanaka, Zhou}, AF metals have no net magnetism
and nevertheless allow transport~\cite{MacDonald, KhaliliAmiri}. This unusual
characteristic makes them highly interesting as spin injectors in spin valves or
magnetic tunnel junctions for future applications~\cite{Iusipova, Wang2, Park}.

From the view of fundamental physics, AF metals are of particular
interest due to the interplay between charge transport, spin, optical, and
magnetization dynamics~\cite{Zhang, Sklenar, Zelezny}. As mentioned above,
antiferromagnetism and metallicity are typically incompatible due to the
significant scattering of conduction electrons by anti-aligned spins~\cite{Takahashi}.
So, most of the identified AF metals have at least one
ferromagnetic nearest-neighbor spin coupling, giving rise to anisotropic
conductivity~\cite{Siddiqui}. Doped manganites $\rm La_{1-x} Sr_x Mn O_3$
exhibit A-type AF metallic ground states for various $x$
values~\cite{Akimoto, Moritomo, Hemberger, Santos}.
A-type AF metallic ground states have also been
reported in $\rm Sr_{2-x}La_xFeMoO_6$ ($x > 1.0$) compounds~\cite{Jana}.
The double-layered $\rm Ca_3Ru_2O_7$ exhibits A-type AF
order below 56~K and a metal--nonmetal transition near 48~K, resulting in an
AF metallic phase in the intermediate temperature range
(48-56~K)~\cite{Yoshida}. $\rm CrN$ shows a first-order magnetic-structural
transition from a paramagnetic (PM) insulator to an AF metallic
phase at Neel temperature $T_N = 286$~K~\cite{Bhobe}, for which AF ordering is
still unclear. $\rm CaCrO_3$ and $\rm SrCrO_3$ are examples
of metallic antiferromagnets (C-type), where hybridized Cr$3d$--O$2p$ states
at the Fermi level play an essential role~\cite{Komarek, Bhobe2, Ortega, Zhang2}. 

As far as we are aware, G-type AF metals are incredibly rare in
stoichiometric compounds, despite the fact that a range of materials have been
discovered to be AF metals, offering a platform to explore this
unusual state. This is because anisotropic conductivity, which can occur in A-type
and C-type antiferromagnets, is almost impossible in G-type antiferromagnets.
But, recently, the A-site-ordered cubic perovskite oxide $\rm LaCu_3Cr_4O_{12}$
was found to display an unusual metallic and G-type AF state
at 225~K~\cite{Saito}. This unique behavior can be attributed to the partial
replacement of Cu sites by La sites in the material. The strong hybridization
between Cu-$3d$, Cr-$3d$, and O-$2p$ orbitals around the Fermi level
then facilitates this unusual electronic structure.

The chemical replacement of B-site magnetic ions in $ABO_3$ perovskites with
nonmagnetic impurities also results in the emergence of G-type AF
metallic states. In certain materials, replacing a magnetic ion disrupts the
existing exchange channels, which dilutes the magnetic interactions within
the material but increases charge transport by changing the quantity of
charge carriers. For example, in $\rm LaFe_{1-x}Mo_xO_3$, where at around
$25\%$ $\rm Mo$ doping~\cite{Phuyal,Jana4}, the AF insulating ground
state of undoped $\rm LaFeO_3$ transitions to an AF metallic phase.
The observed metallicity is attributed to itinerant carriers linked to the $\rm Mo$,
as determined by resonant photoemission spectroscopy. In CuO, a type-II antiferromagnet,
substituting $\rm Li^+$ ions for $\rm Cu^{2+}$ ions also dilutes the magnetic 
lattice and adds charge carriers to the system~\cite{Ghijsen, Yang, Misawa}. The
competing interactions between localized spins and itinerant carriers enhance the
charge transport, lowering resistivity values while maintaining the AF
ground state at low temperatures~\cite{Zheng, Siddiqui2}. On the other hand, $\rm CaMnO_3$,
a G-type AF insulator with $T_N \approx 120$~K~\cite{Wollan, MacChesney}
undergoes a sharp change into a C-type AF metal with as little as
$1\%$ $\rm Nb^{5+}$ substitution at the $\rm Mn$ site~\cite{Xu, Markovich}. The
combination of mixed-valence states, itinerant carriers, and competing magnetic
interactions causes a considerable decrease in resistivity and the suppression of
$T_N$ in many of these transitions.

Can isovalent substitution, when the dopant has the same nominal valence as the host
ions, induce metallicity in an antiferromagnet? A few of the factors that might be
essential in modifying the electronic conductivity and magnetic exchange interactions
under isovalent doping are structural distortions, orbital overlaps, and changes in
the bond angles. A few investigations have been performed to ascertain whether
isovalent substitution can give rise to antiferromagnetism and metallicity at the
same time by affecting the local electronic structure and bandwidths without introducing
additional carriers. It was shown that the $T_N$ and the resistivity both decrease
when $\rm Al^{3+}$ ions replace $\rm Cr^{3+}$ ions in $\rm LaCrO_3$ by a small
percentage~\cite{Silva, Wang3}. Substituting $\rm Ni^{2+}$ with
nonmagnetic elements such as $\rm Mg^{2+}$~\cite{Feng, Das}
and $\rm Ag^{2+}$~\cite{Benseghier} also weakens the AF
order in NiO, reduces the band gap, and decreases the resistivity.

Overall, the inclusion of foreign dopants in the system promotes competitive
magnetic interactions that destabilize the parent spin configurations by triggering
local anisotropies, lattice distortions (from size mismatch), changing the orbital
overlaps, and altering the magnetic exchange pathways. The competing interactions,
commonly referred to as magnetic frustration, due to the nonmagnetic impurities, can
have a significant, often complex, effect on the magnetotransport properties of
correlated materials. For example, in $\rm La_2CuO_4$ (a G-type antiferromagnet),
replacing magnetic $\rm Cu^{2+}$ ions with nonmagnetic $\rm Mg^{2+}$ or $\rm Zn^{2+}$
ions disrupts the spin network and lowers the $T_N$ because of the enhanced magnetic
frustration, even though the magnitude of the ordered $\rm Cu$ moments stays
essentially unchanged at low temperatures~\cite{Chakraborty, Cheong, Wan, Vajk, Liu2}.

In this paper, we investigate the combined effect of site dilution and long-range
interactions on the magnetotransport properties in the context of strongly correlated
electron systems, motivated by the above interesting experimental results, in an effort
to provide a more comprehensive qualitative understanding of the underlying mechanisms
that have not yet been investigated. Our focus is on a $t$-$t'$ ($t$: nearest-neighbor and
$t'$: second-nearest-neighbor hopping amplitudes) based diluted one-band Hubbard model
on a simple cubic lattice to describe the interplay between electronic 
correlations~\cite{Delannoy, Langmann, Raczkowski}, competing interactions~\cite{Lin, Tocchio, Laubach, Jana2}, 
and site dilution~\cite{Ulmke, Fratino, Chakraborty2}. Key phenomena like magnetic phase
transitions~\cite{Staudt, Yu, Gebhard} and metal-insulator transitions~\cite{Kondo, Becca, Werner, Igoshev} 
are particularly well-suited for investigation using the undiluted Hubbard model~\cite{Staudt, Mandal}.
To account for site dilution in our investigation, we set the on-site repulsive Hubbard interaction 
strength ($U$) to zero on a fraction of sites. Tuning the relative strengths of $t$ and
$t'$ for different site dilution concentrations ($x$) in the presence of on-site repulsive Hubbard interaction
allows us to understand how competing interactions affect the magnetotransport phases in
the system.

The $t$-$t'$ competition has significant implications: it modifies the effective bandwidth,
reshapes the Fermi surface, and affects the transport and magnetic properties~\cite{Jana2, Mandal, Seehra}.
Furthermore, the introduction of random site dilution results in spatial inhomogeneities
and disruption of magnetic ordering, offering a controlled way to examine disorder-driven
insulator-metal transitions without affecting the underlying magnetic phases of the
system~\cite{Chakraborty2, Fratino, Delannoy2}. In fact, inhomogeneities due to dilution
may allow percolation to occur through the lattice to alter the transport
properties. As a result, metallic behavior can develop even in
parameter regimes where insulating states are generally preferred in the clean limit.
All things considered, the combined effects of $t'$, $U$, and $x$
produce a very nontrivial phase landscape where subtle alterations in microscopic parameters
can result in qualitatively different behaviors --- from strongly correlated disordered
insulating or metallic states to various types of ordered AF insulators or
metals.

We use a well-established semiclassical Monte Carlo (s-MC)
technique~\cite{Mukherjee, Chakraborty2, Jana2, Halder, Mandal} to analyze magnetotransport
properties. This approach allows us to capture the complex relationships between frustration,
electronic correlations, and disorder-induced effects that help us to systematically
investigate the vast phase space of the system~\cite{Chakraborty2, Mandal}. Our primary
focus is on the nonperturbative interaction regime, where the electronic bandwidth and
the on-site repulsive Hubbard interaction strength are comparable. For correlated materials, this
intermediate coupling regime is especially important because it captures the essential
competition between the kinetic energy and the Coulomb repulsion that controls
many-body ground states and phase transitions~\cite{Chakraborty2, Mandal, Halder}.

This paper is organized as follows: The model Hamiltonian and numerical methods
used in this investigation are introduced in Section~\textbf{\ref{sec_mm}}.
Appendix~\textbf{\ref{derivation_Heff}} contains additional technical information
about the model Hamiltonian. The procedures used to compute various physical
observables relevant to the analysis of magnetotransport properties are described
in Appendix\textbf{~\ref{obs}}. We focus on analyzing the magnetotransport behavior
of the diluted Hubbard model without any long-range hopping for $U \sim$ bandwidth
in Section~\textbf{\ref{sec_tt0}}. Next, the effects of the second-nearest-neighbor
interactions on the magnetic and transport properties are examined in
Section~\textbf{\ref{sec_tt}}. In Section~\textbf{\ref{sec_u}}, we show how the
Hubbard on-site repulsion strength affects the ground state phases to give a
wider perspective of our work. The cumulative impact of the first-, second-, and
third-nearest-neighbor interactions is then examined in Section~\textbf{\ref{sec_ttt}}.
Subsequently, in Section~\textbf{\ref{sec_tt12}}, we investigate how the
sublattice-dependent second-nearest-neighbor hopping can be used to engineer
half-metallic antiferromagnets. Finally, Section~\textbf{\ref{sec_con}} summarizes the key findings of our work.

\section{Model Hamiltonian and Method} \label{sec_mm}

To investigate the magnetotransport properties of diluted magnetic
systems we consider the following form of the $t$-$t'$-$t''$ one-band
Hubbard Hamiltonian~\cite{Jana2, Mandal} on a simple cubic lattice with
periodic boundary condition:
\begin{align} \label{hamiltonian}
H =& -t \!\! \sum_{\left\langle i,j \right\rangle , \sigma} \!\! c_{i,\sigma}^\dagger c_{j,\sigma} -t' \!\!\!\! \sum_{\left\langle \left\langle i,j \right\rangle \right\rangle, \sigma} \!\!\!\! c_{i,\sigma}^\dagger c_{j,\sigma}-t'' \!\!\!\! \sum_{\left\langle \left\langle \left\langle i,j \right\rangle \right\rangle \right\rangle, \sigma} \!\!\!\! c_{i,\sigma}^\dagger c_{j,\sigma} \nonumber \\ 
&+ U \sum_i \left(n_{i,\uparrow} -\frac{1}{2}\right) \left( n_{i,\downarrow} -\frac{1}{2}\right) - \mu \sum_i n_i  \nonumber \\ 
=& H_0 +H_I.
\end{align}
Here, the operator $c_{i,\sigma}$ ($c_{i,\sigma}^\dagger$) denotes the fermion
annihilation (creation) operator at site \( i \) with spin \( \sigma \).
The parameters $t$, $t'$ and $t''$ represent the hopping amplitude between 
nearest-neighbor sites, second-nearest-neighbor sites and third-nearest-neighbor
sites, respectively. The number operator at site $i$ with spin $\sigma$ is given
by $n_{i,\sigma} = c_{i,\sigma}^\dagger c_{i,\sigma}$, while the total number
operator at site $i$ is $n_i = \sum_\sigma n_{i,\sigma}$. The parameter $U$ ($>0$)
represents the on-site repulsive Hubbard interaction strength. The chemical potential
$\mu$ regulates the overall carrier density of the system. The term $H_0$
represents the quadratic part of the model Hamiltonian, while $H_I$ corresponds
to the quartic term. In our setup, we assign $U = 0$ to a fraction $x$ of randomly
selected sites ($0 \leq x < 1$ ) and set a finite $U$ at the remaining sites in
order to take into consideration of the dilution in the system. Consequently,
$H_I = U \sum_k n_{k,\uparrow} n_{k,\downarrow}$, where $k$ runs across only
nonzero $U$ sites, is the quartic term.

To tackle the interacting quartic term $H_I$ in our model Hamiltonian, we apply
the Hubbard-Stratonovich (HS) transformation, a standard method for decoupling
interaction terms. This involves introducing auxiliary fields --- specifically, a
vector field $\mathbf{m}_k$ and a scalar field $\phi_k$ --- at $U \neq 0$ lattice
site $k$. These fields allow us to reformulate the Hamiltonian in a way that makes
it amenable to simulation using the s-MC technique.
In this framework, $\mathbf{m}_k$ interacts with the spin sector 
of the system, while $\phi_k$ is tied to the charge degrees of freedom. 
To simplify the problem further, we assume the auxiliary fields 
are static in imaginary time and treat them as classical variables. 
At the saddle-point level, we relate the scalar field to the average local 
electron density via the condition $ i \phi_k = \frac{U}{2} \langle n_k \rangle $. 
Crucially, auxiliary field $\mathbf{m}_k$ is not uniform; it exhibits 
spatial variation and thermal fluctuations, which are central to 
capturing finite-temperature behavior and emergent phases in our model.
Finally, with the above assumptions, we arrive at the following effective
spin-fermion Hamiltonian (please see details in Appendix~\ref{derivation_Heff}): 
\begin{align}\label{h_eff}
H_{eff} = & -t \!\! \sum_{\left\langle i,j \right\rangle , \sigma} \!\! c_{i,\sigma}^\dagger c_{j,\sigma} -t' \!\!\!\! \sum_{\left\langle \left\langle i,j \right\rangle \right\rangle, \sigma} \!\!\!\! c_{i,\sigma}^\dagger c_{j,\sigma} -t'' \!\!\!\! \sum_{\left\langle \left\langle \left\langle i,j \right\rangle \right\rangle \right\rangle, \sigma} \!\!\!\! c_{i,\sigma}^\dagger c_{j,\sigma}   \nonumber \\ 
 & + \frac{U}{2} \sum_k \left( \left\langle n_k \right\rangle n_k - \mathbf{m}_k.{\boldsymbol{\sigma}}_k \right) + \frac{U}{4} \sum_k \left( \mathbf{m}_k^2 - \left\langle n_k \right\rangle^2 \right) \nonumber \\ 
 & -\frac{U}{2}\sum_k n_k - \mu \sum_i n_i
\end{align}
\noindent
where $i$, $j$ go over all sites, and $k$ is assigned only to sites with finite $U$ values.

We analyze this effective spin-fermion Hamiltonian $H_{\text{eff}}$ using a
diagonalization-based s-MC method at half-filling. First, a given auxiliary field
configuration $\{\mathbf{m}_k\}$ and average electron densities $\{\left\langle n_i \right\rangle\}$ 
are used to diagonalize the Hamiltonian at very high temperatures. The Metropolis algorithm
is implemented to update $\{\mathbf{m}_k\}$ (at each finite $U$ sites), whereas
$\{\langle n_i \rangle\}$ (at all the sites) are self-consistently updated after 10 Monte
Carlo steps. The goal is to use iterative updates that generate equilibrium configurations
of $\left\langle n_i \right\rangle$ and $\mathbf{m}_k$ at each temperature. At each
temperature, the resulting eigenvectors and eigenvalues are used to calculate the expectation
values of observables, which are averaged across several equilibrium configurations. We
measure observables in each of the ten MC sweeps to avoid any self-correlation in the data.
To compute temperature-dependent magnetotransport properties, the temperature is gradually
lowered during simulation. We use the Monte Carlo technique combined with a traveling-cluster
approximation (TCA)~\cite{Kumar} using $4^3$ TCA cluster to address system sizes of $L^3 = 10^3$.
In our s-MC investigation, we use $t = 1$ and express $t'$, $t''$, $U$, and temperature ($T$)
in units of $t$.

\section{Magnetotransport Properties: Diluted Hubbard Model without any Extended Hopping ($t' = t'' = 0$)} \label{sec_tt0}

\begin{figure}[t]
\centering
\includegraphics[width=0.48\textwidth]{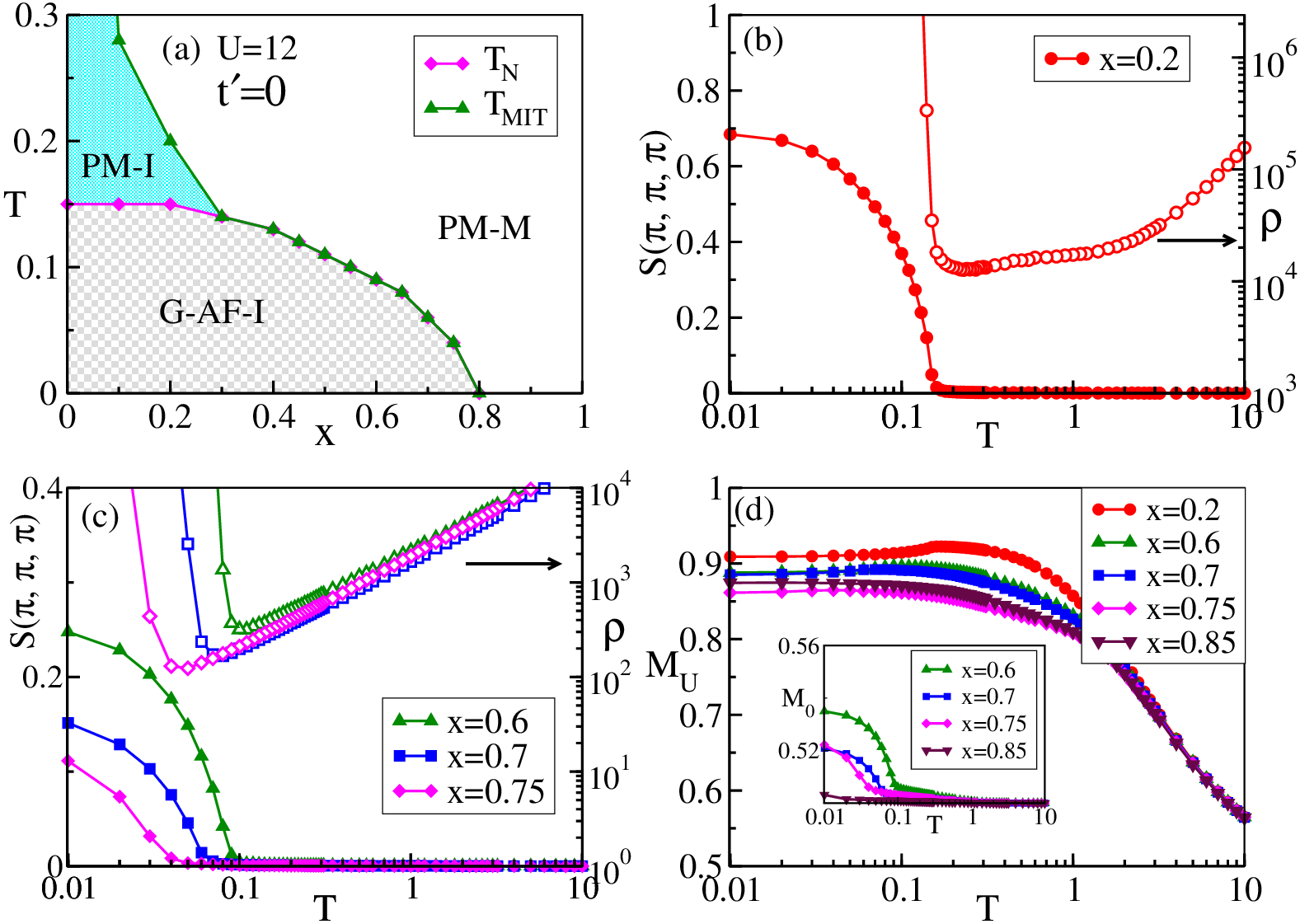}
\caption{
Magnetotransport properties in the absence of long-range interactions
($t' = 0$ and $t'' = 0$): (a) The $x$-$T$ phase diagram for $U = 12$. Here, $x$ denotes
the concentration of $U = 0$ sites (uncorrelated sites), while use $U = 12$ for the
remaining sites. For $x < 0.3$ the $T_{MIT}$
exceeds the $T_N$ due to the appearance of an intervening PM-I
phase between PM-M and G-type AF insulating phases. For $0.3 \leq x<0.8$,
the high-temperature PM-M phase directly transforms into
the G-type AF insulating phase at low temperatures.
The G-type AF order totally diminishes for $x \geq 0.8$, establishing the
PM-M phase even at low temperatures.
The G-type AF insulating phase is labeled as G-AF-I in the phase diagram.
Temperature evolution of the G-type AF structure factor $S(\pi,\pi,\pi)$ [left axis using
solid symbols] and resistivity $\rho$ [right axis using open symbols] are presented
in (b) and (c) for various $x$ values. Here, all sites (both $U = 12$ and $U = 0$)
are used to calculate $S(\pi,\pi,\pi)$ and $\rho$. The $T_{MIT}$ is greater than
the $T_N$ for $x = 0.2$. The $T_N$ and $T_{MIT}$ coincide for $x =$ 0.6, 0.7, and 0.75. 
(d) The average local moments $M_U$ [calculated only using $U = 12$ sites] vs. $T$
for various $x$ values show that the value of $M_U$ decreases as $x$
increases at low temperatures, but remains substantial even at large $x$. The inset,
$M_0$ vs $T$, shows that the induced moment $M_0$ [calculated only using $U = 0$
sites] reduces as $x$ increases and vanishes completely for larger $x$ values.
We use $U = 12$ unless otherwise mentioned in all our calculations.
}
\label{fig01}
\end{figure}

In this section, we examine the magnetotransport properties of the diluted Hubbard model
by only considering nearest-neighbor hopping (i.e., Extended Hopping $t' = t'' = 0$). 
We focus on $U = 12$ (i.e., $U\sim$ bandwidth), which lies within the non-perturbative regime. 
The $x$-$T$ phase diagram is shown for $t' = t'' = 0$ in Fig.~\ref{fig01}(a). The ground state 
of the diluted system exhibits a long-range G-type AF insulating phase at low temperatures for
$x < 0.8$. For $x < 0.3$, a paramagnetic insulating (PM-I) phase separates the paramagnetic
metallic (PM-M) and G-type AF insulating phases. Therefore, the
metal-insulator transition temperature ($T_{MIT}$) surpasses the $T_N$ in this regime. 
However, the $T_{MIT}$ decreases rapidly as dilution increases, whereas $T_N$ stays nearly constant,
indicating that the magnetic order is more robust than the insulating state in this regime.
Consequently, the $T_N$ and the $T_{MIT}$ overlap beyond $x = 0.3$ and decrease
monotonically as dilution (i.e., $x$) increases. Our findings are consistent with previous
results~\cite{Chakraborty2}.

We also present the magnetotransport properties in detail for two separate regimes:
$T_{MIT} > T_N$ and $T_{MIT} = T_N$. The magnetic structure factor $S(\pi,\pi,\pi)$ and
resistivity $\rho$ are plotted with temperature for $x = 0.2$ in Fig.~\ref{fig01}(b). 
For a small dilution, the $T_{MIT}$ exceeds the $T_N$, as previously mentioned. 
On the other hand, the $T_{MIT}$ is comparable to $T_N$ for $x = 0.6$, $0.7$, and $0.75$
[see Fig.~\ref{fig01}(c)]. Crucially, the magnitude of $S(\pi,\pi,\pi)$ at low temperatures
decreases with the increase of $x$. This is because, at low temperatures, both the
number of associated $U = 12$ sites and the amplitude of the moment at these sites
decrease as dilution increases. This decrease in magnetic moment at $U = 12$ sites at
low temperature is clear from the temperature evolution of the average local
moment $M_U$ (calculated using only $U = 12$ sites), as illustrated in Fig.~\ref{fig01}(d).
It is true, however, that even at high levels of dilution, $M_U$ remains significantly
large. Also, for all the $x$ values, as temperature drops, $M_U$ rises and saturates
below $T_N$.

It is noteworthy that long-range magnetic order persists above $x_p = 0.69$, which
is the classical percolation threshold. As seen in Fig.~\ref{fig01}(a), this is due
to the induced moments on the $U = 0$ sites. For $x$ = 0.75 and 0.7, the induced
moment $M_0$ exhibits an upturn around $T_N$ [see the inset of Fig.~\ref{fig01}(d)]. 
But for $x = 0.85$, where long-range magnetic ordering ceases to exist,
the induced moments remain very small even at very low temperatures. This indicates a
one-to-one correspondence between the induced moment at uncorrelated ($U = 0$)
sites and long-range G-type AF order above the percolation threshold.

\begin{figure}[t]
\centering
\includegraphics[width=0.48\textwidth]{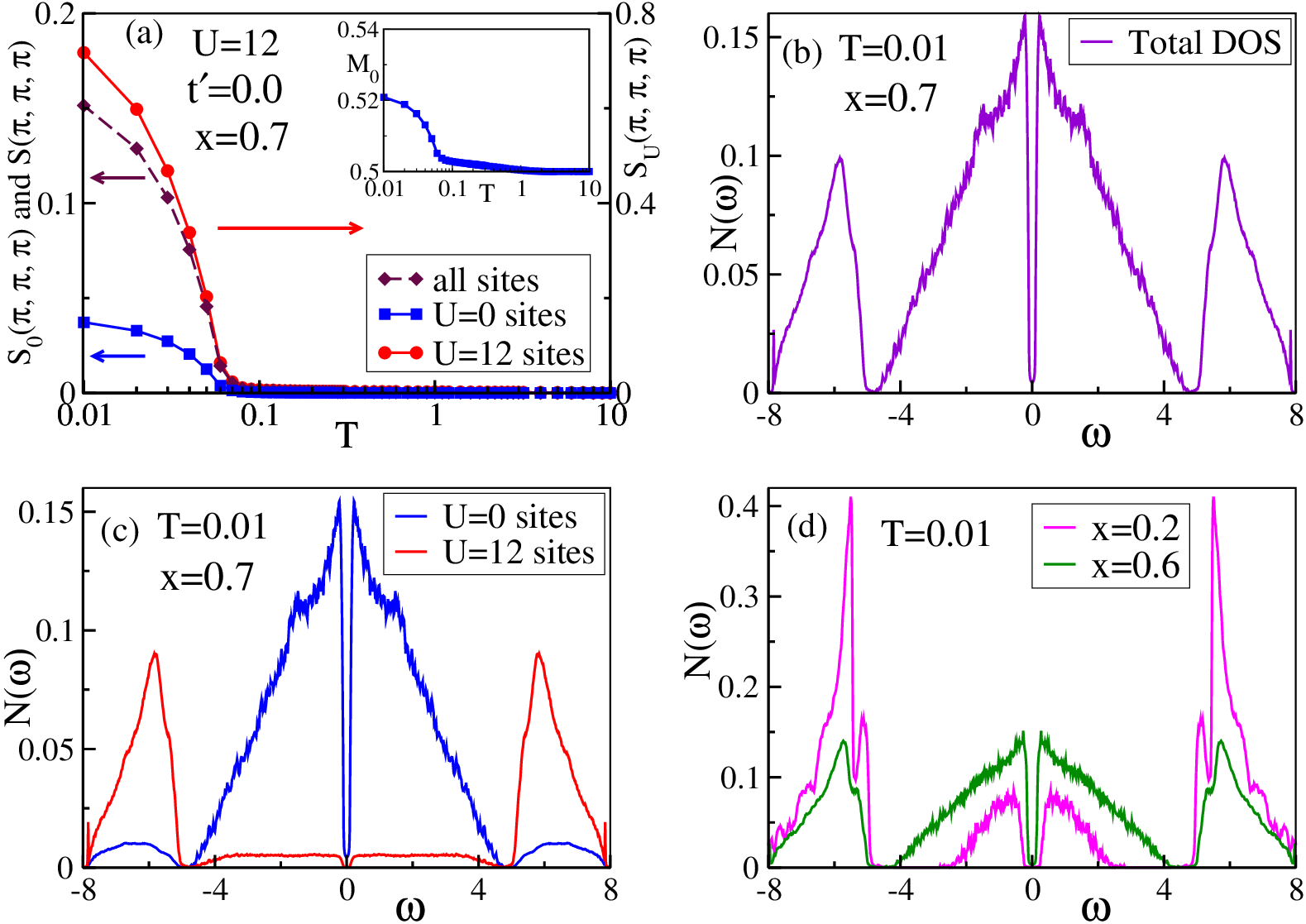}
\caption{
Magnetotransport properties for $x = 0.7$ in the absence of long-range
interactions ($t' = 0$ and $t'' = 0$): (a) Temperature evolution of magnetic structure
factors [$S_U(\pi,\pi,\pi)$ is determined using only $U = 12$ sites, whereas
$S_0(\pi,\pi,\pi)$ is calculated using only $U = 0$ sites] are plotted. Additionally,
the dotted line displayed $S(\pi,\pi,\pi)$, which is calculated utilizing all the
sites. For both the $U = 12$ and $U = 0$ sites, the AF transition
temperature ($T_N$) coincides. The $M_0$ display in the inset also transitions at the
$T_N$, indicating the one-to-one connection between the onset of the magnetic ordering
in the system and the induced magnetic moment at $U = 0$ sites.
(b) The total density of states (DOS) exhibits two pairs of lobes at low temperature
($T = 0.01$). One pair forms around $\sim \pm U/2$, while another forms at two sides
of the Fermi level ($\omega = 0$). Fermi energy is set to zero ($\omega = 0$).
(c) $U$-dependent DOS shows that the pair of lobes that are close to $\omega = 0$
are mainly formed by the $U = 0$ sites.
(d) DOS are compared for $x = 0.2$ and $0.6$ to show that the gap in DOS at the
Fermi level decreases as $x$ increases.
}
\label{fig02}
\end{figure}

Next, we present the $U$-dependent magnetic structure factors (i.e., separately for
$U = 12$ sites and $U = 0$ sites) for $x = 0.7$ in Fig.~\ref{fig02}(a) to further
demonstrate the one-to-one correspondence between the magnetic ordering arising
among $U = 12$ and $U = 0$ sites. The $T_N$ for both types of sites (i.e., $U = 12$
and $U = 0$) is the same, which aligns with the overall system. The temperature
evolution of $S(\pi,\pi,\pi)$ is represented using a dotted line in the same
figure. As seen in the inset of Fig.~\ref{fig02}(a), the induced magnetic moments
also increase at the same temperature. This shows a close connection between
the magnetic ordering of the $U = 12$ and $U = 0$ sites of the system.

What role do these induced magnetic moments at $U = 0$ sites play in maintaining
the insulating properties of the undiluted system? To answer this, we plot the
DOS in Fig.~\ref{fig02}(b) at low temperature ($T = 0.01$).
The DOS clearly shows two pairs of sub-bands, which could be a result of the two
types of sites ($U = 12$ and $U = 0$ sites) that we have taken into consideration.
One pair of sub-bands forms around $\pm U/2$, similar to the Mott lobes in
undiluted systems, while the other pair is located at the two sides of the Fermi
level (Fermi energy is set at zero; $\omega = 0$ unless otherwise specified),
leaving a small gap between them. 

For further analysis of the sub-bands, we plot $U$-dependent DOS in
Fig.~\ref{fig02}(c). The sub-bands at $\pm U/2$ are primarily caused by the $U = 12$
sites and resemble to that of the Mott lobes. The new pair of sub-bands near the Fermi
level is caused by the $U = 0$ sites. Therefore, the $U = 0$ sites form the impurity band,
and the induced magnetic moments in the $U = 0$ sites trigger the gap at the Fermi
level. The insulating nature of the system arises due to this small gap. The gap in the DOS
around the Fermi level gradually decreases with the increase of $x$ values [see Fig.~\ref{fig02}(d)].
It is in good agreement with the fact that the system becomes a weak insulator
as the induced magnetic moments decrease with the increase of $x$.

Overall, as shown in Fig.~\ref{fig01}(a), our calculations show that the $T_{MIT}$
and the $T_N$ decrease with the increase of $x$ values. When the dilution is sufficiently large
($x \geq 0.8$), the system turns out to be in a PM-M state. In this situation,
the gap in DOS disappears, resulting in a metallic ground state, but without any
long-range AF ordering. We would like to pose one important
question at this point: is it feasible to close the gap in the DOS before the
long-range AF order breaks down in a certain dilution range?
In some intriguing cases, dilution can lead to a percolation-like transition
from an insulating state to a metallic state, maintaining the AF state,
highlighting the importance of percolating pathways within the material.
Will long-range magnetic interaction facilitate the achieving of this AF metallic state?
To answer this, we examine the combined effects of $t'$ and $x$ in detail in the next section.

\section{Antiferromagnetic Metal: Role of Second-Nearest-Neighbor Hopping ($t' \ne 0$)} \label{sec_tt}

\begin{figure}[t]
\centering
\includegraphics[width=0.48\textwidth]{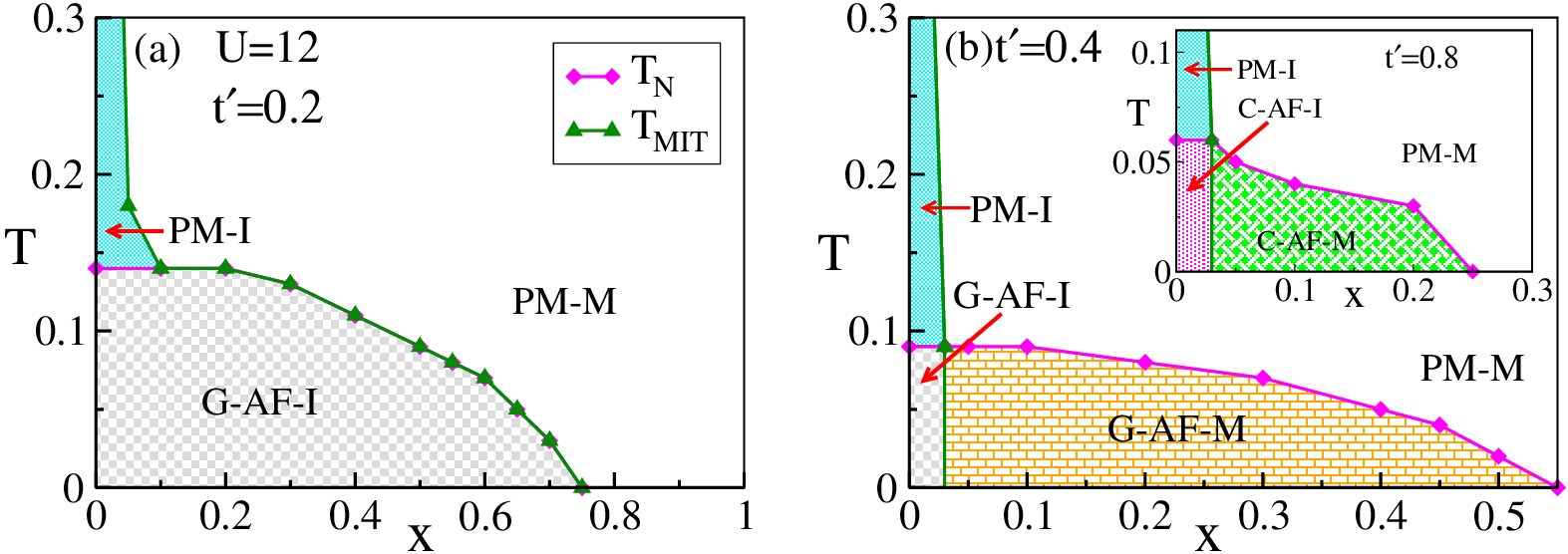}
\caption{The $x$-$T$ phase diagrams at various $t'$ values:
(a) $t' = 0.2$, (b) $t' = 0.4$ [inset: $t' = 0.8$].
(a) The $T_{MIT}$ exceeds the $T_N$ for $x < 0.1$ whereas the $T_{MIT}$ matches well
with the $T_N$ for $x \geq 0.1$. The G-type AF order disappears for $x \geq 0.75$, and
the transition temperatures reduce for all $x$ in comparison to the $t' = 0$ case
[see Fig.~\ref{fig01}(a)].
(b) At $t' = 0.4$, the $T_{MIT}$ as one decreases the temperature remains confined to
small $x$ values ($x < 0.03$). Notably, the G-type AF metallic phase appears for
$0.03 < x < 0.55$ at low temperatures. Magnetic ordering vanishes beyond $x = 0.55$.
When compared to the $t' = 0$ case, the suppression of the $T_N$ and the $T_{MIT}$ is more
noticeable for $t' = 0.4$ than for $t' = 0.2$. For $t' = 0.8$, the inset shows a C-type
AF insulating order close to $x = 0$ (undiluted). So, as $t'$ increases, the G-type AF insulating
phase changes to a C-type AF insulating phase for the nearly undiluted
case ($x \lesssim 0.03$). This C-type AF insulating phase changes into the C-type AF
metallic state when $x$ increases beyond 0.03. 
G-AF-M, C-AF-M, and C-AF-I indicate the G-type metallic, C-type metallic, 
and C-type AF insulating phases, respectively.
Legends are the same in panels (a) and (b).
}
\label{fig03}
\end{figure}

We display the $x$-$T$ phase diagram for $t' = 0.2$ at $U = 12$ for analyzing the effect
of long-range interactions in Fig.~\ref{fig03}(a). The qualitative properties of the phase
diagram remain the same as in the $t' = 0$ case. Specifically, the long-range G-type AF ground
state at $x = 0.75$, which was achievable when $t' = 0$ [see Fig.~\ref{fig01}(a)], is 
absent for $t' = 0.2$. For $t' = 0.2$ the $T_N$ and the $T_{MIT}$ coincide for
$0.1 \le x <0.75$. Below $x = 0.1$, the $T_{MIT}$ surpasses the $T_N$ due to
the presence of the PM-I phase above the magnetic transition.
In general, when $t'$ rises to $0.2$, both $T_N$ and $T_{MIT}$ decrease across all $x$
values in comparison to the $t' = 0$ scenario. This decrease is more noticeable
at larger $x$ values, which correspond to larger levels of dilution in the system.

Notably, as $t'$ grows to 0.4, a specific region of the phase diagram hosts an
AF metallic phase [see Fig.~\ref{fig03}(b)]. In contrast to $t' = 0$ and
0.2 values, an antiferromagnetically ordered state emerges at only smaller range of $x$
values (i.e., for $x < 0.55$). It is also evident that, in comparison to lower $t'$s,
the $T_N$ substantially decreases for all $x$ values. The phase diagram is therefore
dominated by the PM-M regions. Surprisingly, however, most of the AF
states ($0.03 < x < 0.55$) turn out to be metallic, resulting in the intriguing G-type
AF metallic ground state. For $x < 0.03$, the G-type AF insulating phase
remains the ground state, similar to the case where $t' = 0$. So, interestingly, the $t'$
that partially inhibits the G-type AF order facilitates a frustrated-assisted metallic
state in the diluted systems.

As $t'$ increases even more, the magnetic ground state undergoes another significant
change. The G-type AF phase that occurs for lower $t'$ values is completely suppressed
for $t' = 0.8$. Instead, a C-type AF phase appears below $x = 0.25$, as shown in the
inset of Fig.~\ref{fig03}(b). Remarkably, the C-type AF state manifests metallicity
for $0.03 < x <0.25$. On the other hand, similar to the undiluted situation, the
resulting C-type AF phase maintains its insulating properties at low temperatures
for very small dilution regimes ($x < 0.03$). Therefore, the metal-insulator
transition occurs below $x \sim 0.03$. In brief, the system stabilizes into a G-type
AF metallic ground state for moderate values of $t' \sim 0.4$ and intermediate and
small dilution regimes ($0.03 < x < 0.55$), whereas a C-type AF metallic ground
state arises for higher $t' \sim 0.8$ in the small dilution ($0.03 < x < 0.25$) limit.

\begin{figure}[t]
\centering
\includegraphics[width=0.48\textwidth]{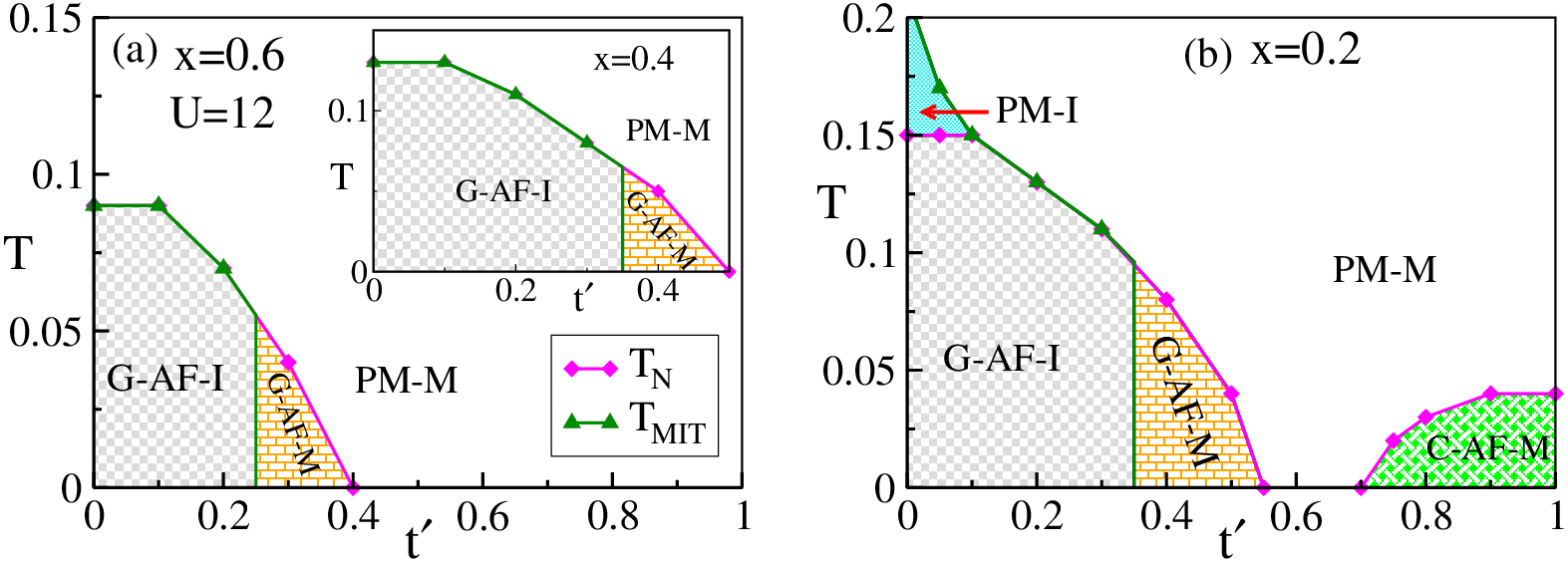}
\caption{The $t'$-$T$ phase diagrams at various dilutions level:
(a) $x = 0.6$ [inset: $x =$ 0.4] and (b) $x = 0.2$. (a) For $x = 0.6$, the $T_N$
and the $T_{MIT}$ coincide for $t' < 0.25$, showing a direct transition as the
temperature decreases from the PM-M phase to the G-type AF
insulating phase. Importantly, the G-type AF metallic phase stabilizes at low
temperatures beyond $t' \approx 0.25$. But, after $t' = 0.4$, the AF ordering
disappears. In the inset, the G-type AF order persists until $t' = 0.5$, as the
dilution decreases to $x = 0.4$. At the same time, the $T_N$ and the $T_{MIT}$
increases for all $t'$, while retaining qualitative features similar to that
of $x = 0.6$. (b) For small dilution ($x = 0.2$), the $T_{MIT}$ turns out to
be greater than the $T_N$ for $t' < 0.1$. Then, the $T_{MIT}$ and the $T_N$
coincide for a short range of $t'$ ($0.1 \le t' < 0.35$). The system maintains
metallicity for $t' > 0.35$ at low temperatures, stabilizing the G-type
AF metallic phase for intermediate $t'$ ($0.35 < t' < 0.55$), and 
the C-type AF metallic phase for large $t' > 0.7$. 
Legends are same in panels (a) and (b).
}
\label{fig04}
\end{figure}

Now we comprehensively explore the $t'$-$T$ phase diagrams for three different $x$
values in order to elaborate on the variation of transition temperatures of the
resulting magnetic states. We begin with a relatively larger $x$ ($= 0.6$, slightly
smaller than the typical percolation limit $x_p = 0.69$), where the system transitions
from a G-type AF insulating phase to a G-type AF metallic phase to PM metallic phase
[see Fig.~\ref{fig04}(a)] with an enhancement of $t'$. The G-type AF insulating state persists
at low temperatures up to $t' \approx 0.25$, and in this $t'$ range, $T_N$ decreases with
$t'$. Beyond that, the G-type AF metallic state becomes stable within a narrow $t'$
window. The system maintains the PM-M state even at low temperatures for
$t' \ge 0.4$ region. This was expected as the large second-nearest neighbor hopping
is likely to disrupt long-range magnetic ordering through competing interactions.
In comparison to $x = 0.6$, the $T_N$ of the G-type AF phase is
somewhat enhanced for $x = 0.4$; however, the qualitative characteristics of the
$t'$-$T$ phase diagram remain unaltered [see the inset of Fig.~\ref{fig04}(a)].

For small $t'$ values, the $T_{MIT}$ and the $T_N$ overlap in the two previously
mentioned situations ($x = 0.6$ and $0.4$). But, at $x = 0.2$, where dilution is
smaller compared to $x = 0.4$ and $0.6$, the $T_{MIT}$ exceeds $T_N$ at $t' = 0$
[see Fig.~\ref{fig04}(b)] as previously stated. When small $t'$ value is included
(e.g., $0.1 \le t' < 0.35$), the $T_{MIT}$ converges with the $T_N$, analogous to
larger $x$ values. As the temperature decreases in this $t'$ range, a direct
transition from a PM-M phase to a G-type AF insulating phase is observed without
the occurrence of the PM-I phase in between. It is also evident that,
like $x = 0.4$ and $0.6$ cases, the $T_N$ diminishes with $t'$. Interestingly, the
system transitions to a metallic state by increasing $t'$ to $0.35$, while
maintaining the G-type AF order, allowing for the stabilization of the AF
metallic state. This AF metallic state is stable across a slightly wider
range of $t'$ than that of $x = 0.6$ and importantly has larger $T_N$ values.
At $t' = 0.55$, the G-type AF order finally disappears, and at low temperatures,
a PM-M state persists as the ground state. But, unexpectedly, unlike the larger
$x$ values, the ground state becomes C-type AF metallic state for large
$t'$ ($>0.7$).

\begin{figure}[t]
\centering
\includegraphics[width=0.48\textwidth]{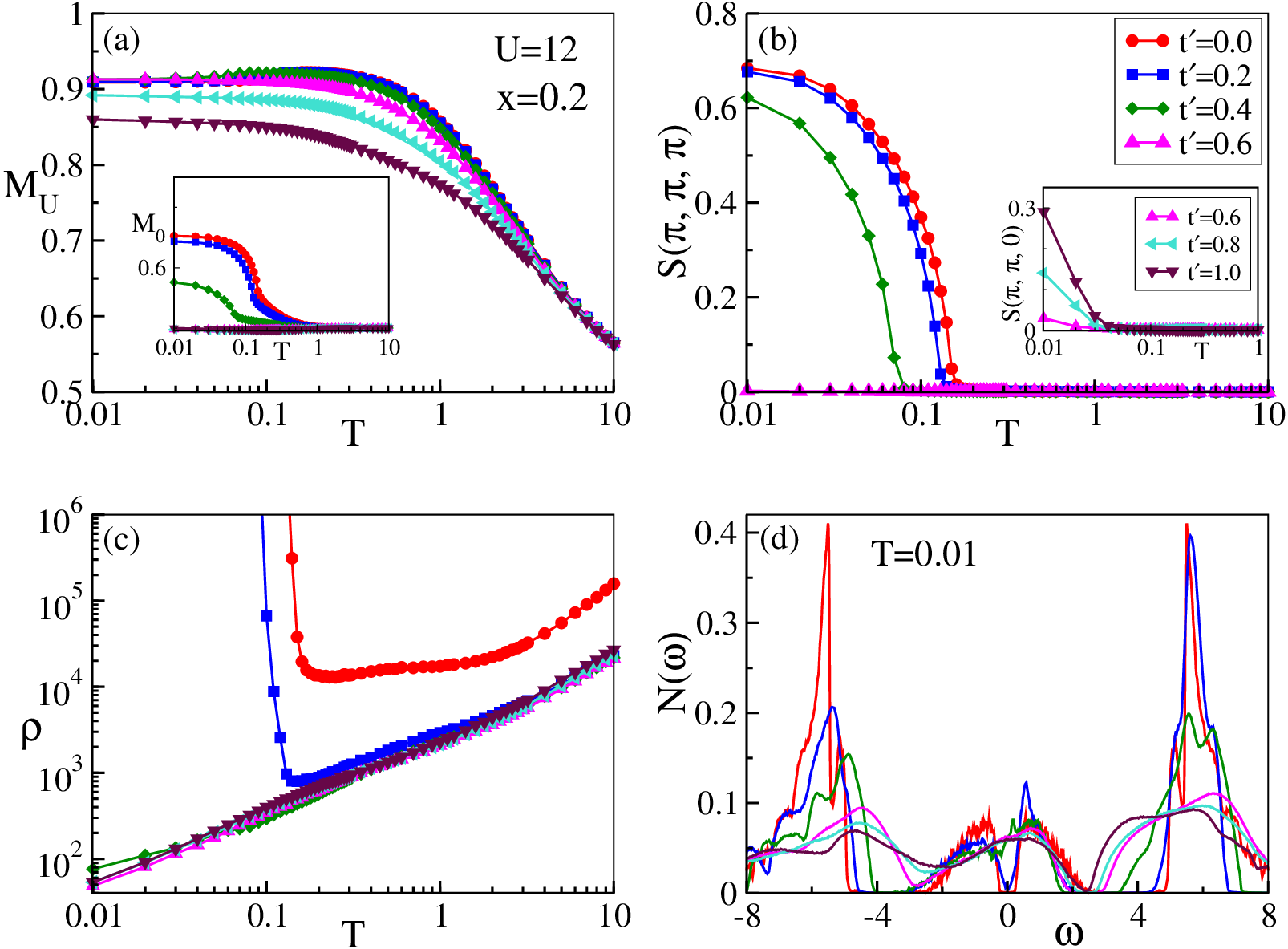}
\caption{Physical observables for $x = 0.2$ with various $t'$:
(a) As $t'$ increases from $0$ to $0.6$, the value of $M_U$ at low temperatures
remains nearly constant, whereas it decreases at $t' = 0.8$ and $1$. At the same
time, $M_0$ reduces with $t'$ and disappears at $t' = 0.6$ and higher (see the inset).
(b) For $t' = 0$, $0.2$, $0.4$, and $0.6$, the temperature evolution of the magnetic
structure factor corresponds to G-type AF order; in contrast, the inset displays
the C-type AF order obtained for $t' = 0.6$, $0.8$, and $1$. For G-type AF ordering, the
$T_N$ decreases as $t'$ increases and disappears for $t' = 0.6$. For $t' > 0.6$, a
C-type AF ordering appears, and $T_N$ marginally increases as $t'$ increases from
$0.8$ to $1$. (c) Resistivity ($\rho$) vs temperature plots for  various $t'$
($= 0$, $0.2$, $0.4$, $0.6$, $0.8$, and $1$) show that the $T_{MIT}$ decreases as $t'$
increases and vanishes for and above $0.4$ --- indicating that the ground state is
metallic. (d) DOS at $T = 0.01$ displays a gap around Fermi level ($\omega = 0$)
at $t' = 0$ and $0.2$, demonstrating an insulating ground state that corroborates
the results in (c). On the other hand, for $t' = 0.4$, $0.6$, $0.8$, and $1$, a finite
value at the Fermi level in the DOS confirms the metallic ground state. Legends
are same in panels (a)$-$(d).
}
\label{fig05}
\end{figure}

To provide a detailed explanation of the effect of $t'$, we now present several
physical quantities for $x = 0.2$ at various $t'$. At low temperatures, as illustrated
in Fig.~\ref{fig05}(a), the saturation value of $M_U$ initially stays relatively
constant as $t'$ increases. This demonstrates that until $t' = 0.6$, the impact of
$t'$ on the magnetic moment is not noticeable. After that $t'$, the saturation value
of $M_U$ decreases, although remains considerably large. The temperature evolution of
the $M_0$ plot [see the inset of Fig.~\ref{fig05}(a)] indicates that the induced
magnetic moments in $U = 0$ sites also disappear beyond the same $t'$ value.
Are the magnetic properties of the whole system related to the induced
magnetic moments at $U=0$ sites? In order to address this, we present the $S(\pi,\pi,\pi)$
vs temperature corresponding to G-type AF order in Fig.~\ref{fig05}(b) for small
to intermediate $t'$ ($t' = 0$, $0.2$, $0.4$, and $0.6$) values. The $T_N$ for G-type
AF order decreases as $t'$ increases and finally disappears for $t' = 0.6$. Please
keep in mind that at $t' = 0.6$, the induced magnetic moments also disappear altogether.
This clearly shows that there is a strong correlation between the stabilization of
G-type AF order at low temperatures and the induction of magnetic moments at $U = 0$
sites. Please note that the magnetic moments are not induced at $U = 0$ sites when
PM-M and C-type AF metallic phases are stabilized, i.e., at larger $t'$. The magnetic
structure factors for C-type AF order at large $t'$ ($0.6$, $0.8$, and $1$) are plotted
in the inset of Fig.~\ref{fig05}(b). Thus, it is evident that the C-type AF order
appears when $t$ and $t'$ are comparable.

We display resistivity ($\rho$) vs. temperature for different $t'$ at $x = 0.2$ in
Fig.~\ref{fig05}(c) to demonstrate the systematic nature of the metal-insulator
transition at low temperatures. As $t'$ grows, the system transitions from G-type AF
insulating to G-type AF metallic, PM metallic, and finally C-type AF metallic at low
temperatures. Initially, only the $T_{MIT}$ decreases as $t'$ increases, but the system
continues to be insulating state at low temperatures. For moderate values of $t'$,
the $T_{MIT}$ eventually vanishes, and the
system enters a low-temperature metallic phase characterized by G-type AF
order. As $t'$ increases further, the system transitions from PM to C-type AF phase
while maintaining its metallic property. We demonstrate the DOS at
$T = 0.01$ for different $t'$ in Fig.~\ref{fig05}(d) to corroborate with the ground
state transport properties. Total DOS displays a distinct gap at the Fermi level
($\omega=0$) for $t' = 0$ and $0.2$, which corresponds well to the insulating ground
states for these $t'$ values. On the other hand, a finite value of DOS arises across
the Fermi level for $t' \ge 0.4$. This indicates that the ground state
shifts toward the metallic regime, which is consistent with the resistivity data.

\begin{figure}[t]
\centering
\includegraphics[width=0.48\textwidth]{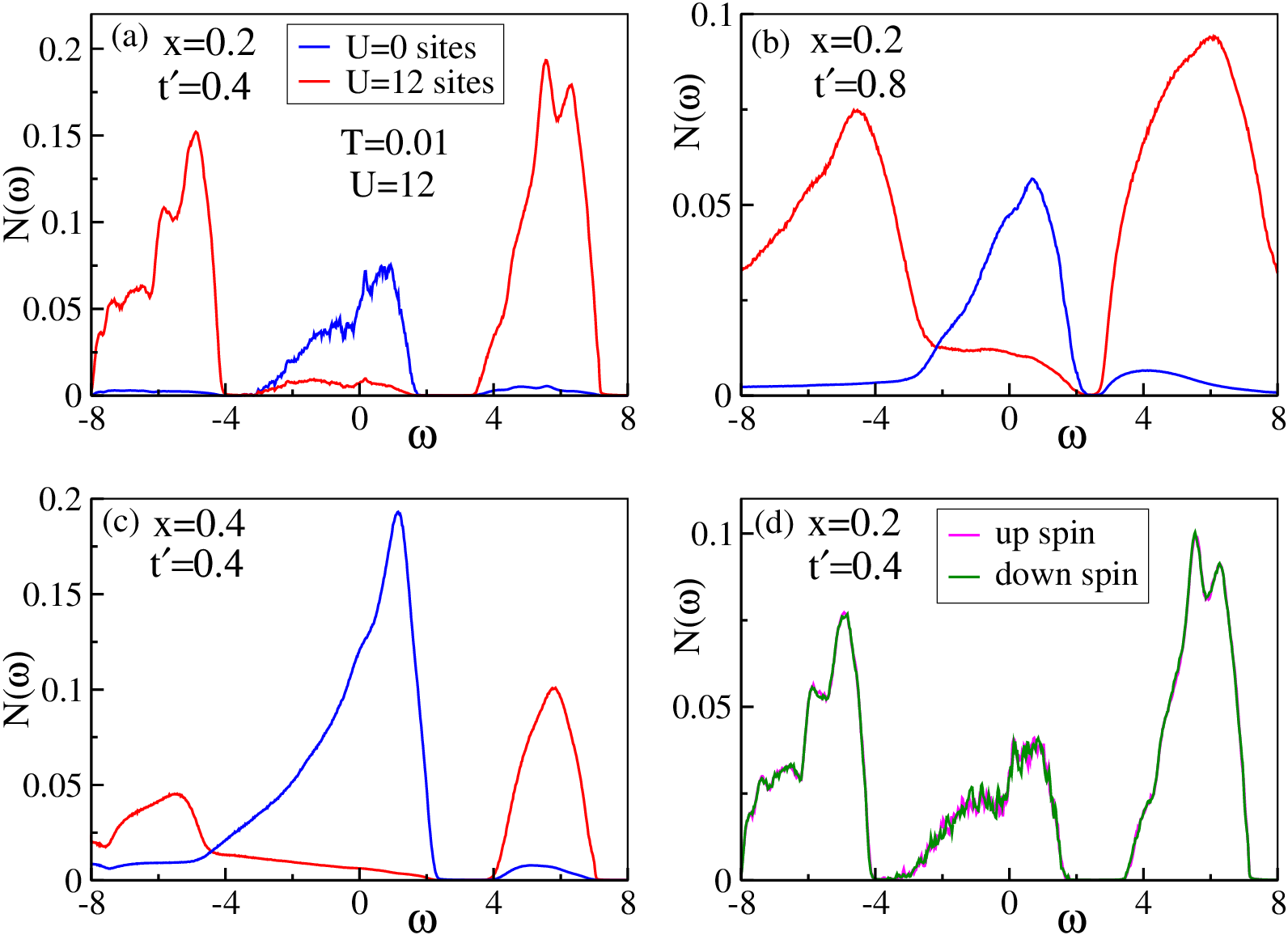}
\caption{$U$-dependent DOS for different combinations of $x$ and $t'$ at low
temperature ($T =$ 0.01):  (a) $x = 0.2$, $t' = 0.4$, (b) $x = 0.2$, $t' = 0.8$, 
(c) $x = 0.4$, $t' = 0.4$.
A noticeable peak appears at the Fermi level ($\omega = 0$) in each plot due to the
$U = 0$ sites, suggesting that the $U = 0$ sites are the source of
the metallic behavior. The prominent DOS near $\pm U/2$ is contributed by the
correlated ($U = 12$) sites. The legends in panels (a)$-$(c) are the same.
(d) For $x = 0.2$ and $t' = 0.4$, the spin-resolved DOS depicts that the system
is not spin polarized.
}
\label{fig06}
\end{figure}

The existence of G-type AF metallic phases at intermediate $t'$ values is very clear
from the Figs.~\ref{fig03} and~\ref{fig05}. For such $t'$ values, e.g., $t' = 0.4$, it
is also apparent that the magnetic moments are induced in $U = 0$ sites. What stabilizes
the metallic state in the AF phase? Is this due to the percolative path
formed by $U = 0$ sites in the system? At this point, it is important to note that the
induced moment at $U = 0$ sites for the AF metallic state is significantly less than that
of the AF insulating state [see the inset of Fig.~\ref{fig05}(a)]. In particular, the
induced magnetic moment at the $U = 0$ site decreases significantly when we shift from
$t' = 0.2$ to $0.4$ at low temperatures (where the system shifts from AF insulating state to
AF metallic state), in contrast to the reduction that occurs when $t'$ rises from $0$ to
$0.2$. So, to investigate the possibility of percolation that may arise from $U = 0$ sites,
we plot the $U$-dependent DOS in Fig.~\ref{fig06}(a) for $x = 0.2$ and $t' = 0.4$. The
primary source of the metallic ground state is the finite DOS at the Fermi level, which
is mostly contributed by $U = 0$ sites. The pair of sub-bands at $\pm U/2$ is due
to the $U = 12$ sites, which is similar to the situation that was previously discussed
in Fig.~\ref{fig02} for the undiluted case. However, the symmetric character of the
sub-bands around the Fermi level is not preserved in the presence of $t'$. Overall, our
calculations strongly suggest that the metallic ground state originates from the
percolating routes formed by uncorrelated ($U = 0$) sites under the influence of
second-nearest-neighbor hopping, which was not feasible in the case of $t' = 0$.

A similar situation is depicted in the $U$-dependent DOS for $t' = 0.8$, where the
C-type AF metallic state is stabilized as the ground state, as shown
in Fig.~\ref{fig06}(b). Additionally, in Fig.~\ref{fig06}(c) we illustrate the
$U$-dependent DOS for $x = 0.4$ at $t' = 0.4$, where the ground state is the G-type
AF metallic state. Here, $U = 0$ sites also assist in maintaining a finite DOS
at the Fermi level that gives rise to percolative conduction in the system.

The next question is whether these metallic states are half-metallic, i.e.,
spin-polarized. To answer this, the spin-polarization of the DOS at the Fermi
level needs to be examined. So, we plot the spin-resolved DOS, shown in
Fig.~\ref{fig06}(d) for $x = 0.2$ at $t' = 0.4$ where ground state shows
metallicity. The DOS for the up and down spin channels overlap, suggesting
that the system is not spin-polarized. In one of the following sections, the
prospect of developing half-metallic antiferromagnets---that is, systems
with complete spin polarization at the Fermi level---will be investigated.

\begin{figure}[t]
\centering
\includegraphics[width=0.48\textwidth]{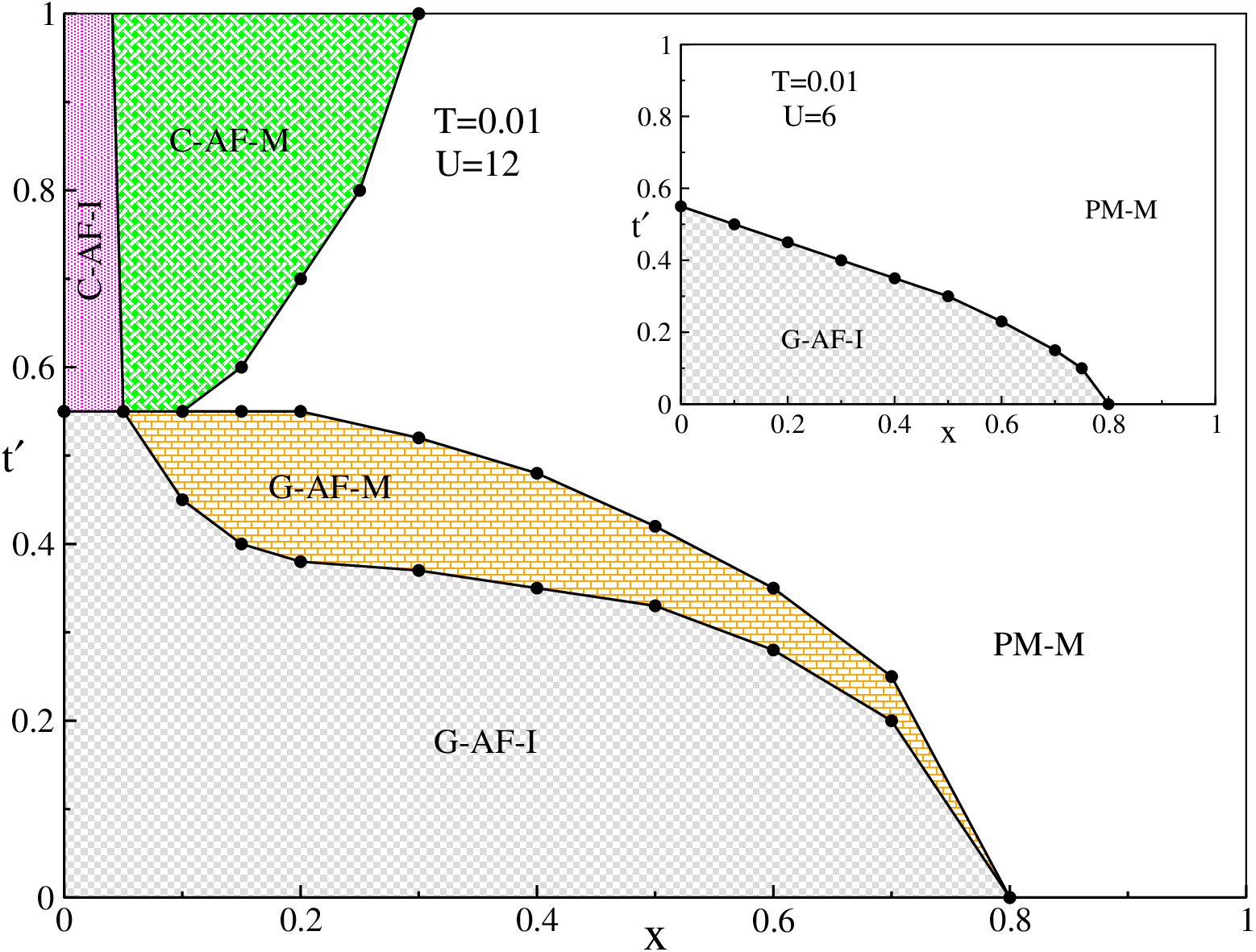}
\caption{The $x$-$t'$ phase diagram for $U = 12$ (main panel) and $U = 6$ (inset) at
$T = 0.01$. For $U = 12$, as $t'$ increases, the system shows clear multiple phase
transitions, particularly at small $x$ values. The transition from G-type AF
insulating to G-type AF metallic phase occurs at small to moderate
$t'$ values within the range $0.05 < x < 0.8$. A C-type AF phase appears as
the ground state for $x < 0.3$ and large $t'$ zone. The system enters into an
insulating state for all $t'$ as $x$ gets closer to $0$, which denotes the low
dilution. Please note that the combination of large $x$ and $t'$ values suppresses
the magnetic ordering. The AF metallic regions are absent for $U = 6$ (see the inset).
}
\label{fig07}
\end{figure}

Now, to illustrate comprehensive results from the combined effect of dilution and second-nearest-neighbor hopping,
we provide the $x$-$t'$ phase diagram for $U = 12$ at $T = 0.01$ in Fig.~\ref{fig07}.
For $0.05 < x < 0.8$, the G-type AF insulating ground state transitions to the G-type AF
metallic phase as $t'$ increases. The G-type AF ordering diminishes for large $t'$, and the
ground state for $x \ge 0.3$ evolves into PM-M. In other words, with the increase in $t'$,
the G-type AF state shifts to the PM-M phase through the G-type AF metallic phase.
The C-type AF metallic phase, on the other hand, appears as the ground state for $x < 0.3$
and large $t'$, with the exception of $x$ values that are near $x = 0$. As $x$ approaches
$0$ (almost undiluted), the ground state transitions to C-type AF insulating from
C-type AF metallic state. Thus, the influence of dilution is minimal on the ground state
properties for all $t'$ values when $x < 0.05$.

For $U = 6$, we illustrate the $x$-$t'$ phase diagram in the inset of Fig.~\ref{fig07}.
At $t' = 0$, the system orders antiferromagnetically for $x < 0.8$. For these $x$ values
($x < 0.8$), the G-type AF insulating phase directly changes to the PM-M phase when
$t'$ increases. So, the AF (G- and C-type) metallic ground states in diluted
systems cease to exist for smaller $U$ values, as established at the $U \sim$ bandwidth
limit. We examine the effects of the $U$ values on the ground state properties in the
next section.

\section{Stability of Antiferromagnetic metallic regime: Variation of On-site Repulsive Hubbard Interaction Strength ($U$)}	\label{sec_u}

\begin{figure}[t]
\centering
\includegraphics[width=0.48\textwidth]{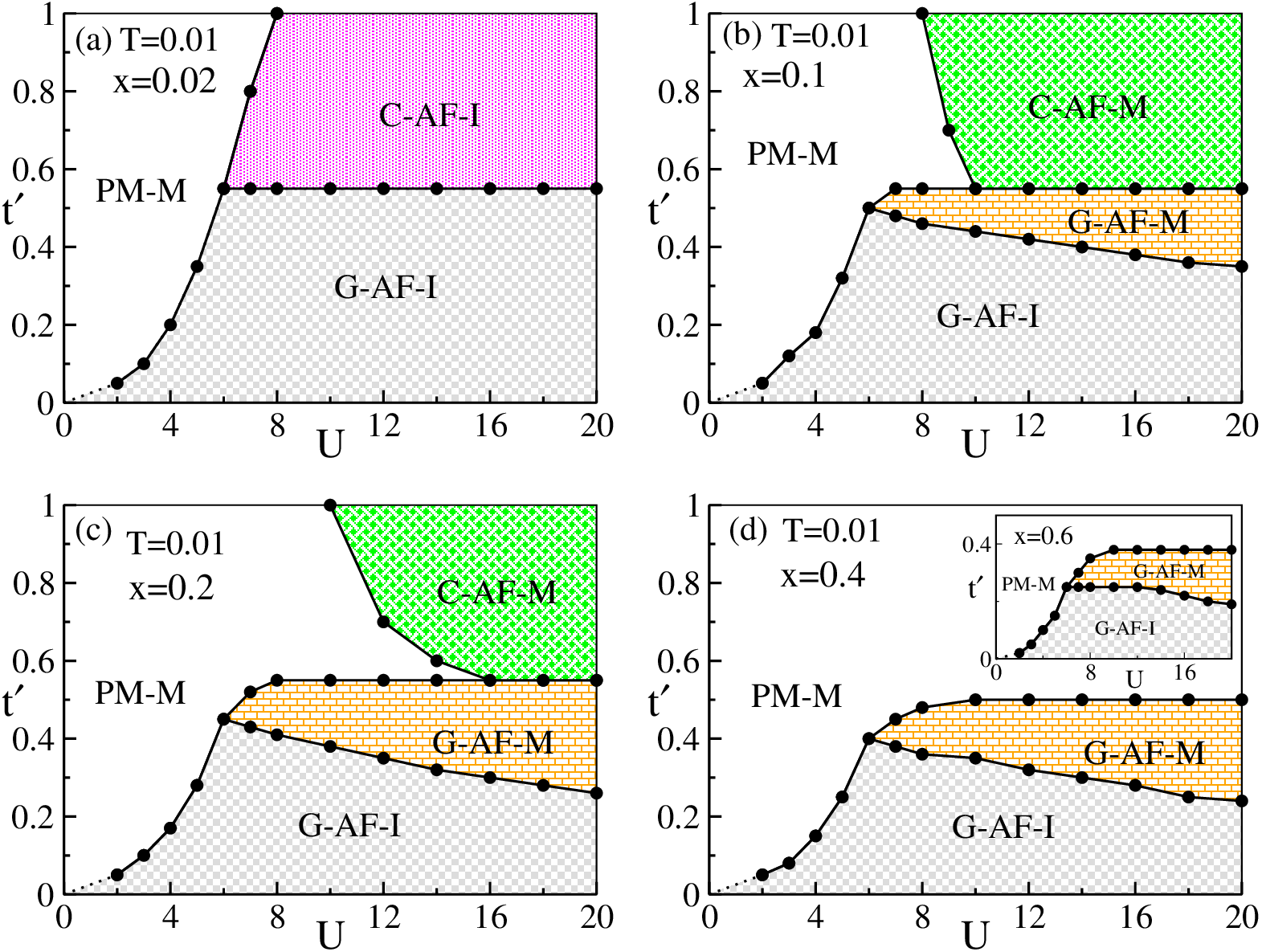}
\caption{The $U$-$t'$ phase diagrams at various dilutions: (a) $x = 0.02$,
(b) $x = 0.1$, (c) $x = 0.2$, and (d) $x = 0.4$ [inset: $x = 0.6$].
In all cases, the G-type AF insulating ground state transitions to the PM-M phase
as $t'$ increases for $U \le 6$. It takes a smaller $t'$ to destabilize the AF order
for a smaller $U$ value. For intermediate and large $U$ values, ground state magnetic
phases encounter a major shift as $t'$ increases. The G-type AF insulating phase transitions to the
C-type AF insulating phase at $x = 0.02$, with an enhancement of $t'$ for
$U > 6$.  When $x = 0.1$, the G-type AF metallic phase, which is stable
at moderate $t'$, appears in between the PM-M phase and the G-type AF insulating
phase for $6 < U < 10$. Upon increasing $t'$ for $U \ge 10$, the G-type AF metallic
phase transitions directly to the C-type AF metallic phase. The phase diagram for
$x = 0.2$ is qualitatively similar to $x = 0.1$. Notably, the G-type AF metallic
region grows larger as we increase $x$ from $0.1$ to $0.2$. The large $t'$ assisted
C-type AF order is suppressed with further dilution ($x = 0.4$ and $0.6$). 
Additionally, it is evident that the G-type AF metallic phase narrows with increasing 
$x$ as we go beyond $x = 0.2$ and eventually disappears for very large $x$.
}
\label{fig08}
\end{figure}

We now investigate the $U$-$t'$ phase diagram for a few different $x$ values to
determine the range of $U$ at which the system transitions to an AF metallic state
in the presence of $t'$. This will also help in understanding how $t'$ affects the
ground state as dilution increases for different values of $U$. Similar to an undiluted
system, the ground state stays G-type AF insulating for all $U$ values at $t' = 0$ for
a given small dilution limit (say at $x = 0.02$) as shown in Fig.~\ref{fig08}(a). In
fact, for $x = 0.02$ (i.e., very low dilution), our calculations show that the ground
state $U$-$t'$ phase diagram is comparable to the undiluted case~\cite{Mandal} where
all of the resulting ordered states (G- and C-type AF states) remain insulating, and
hence it is clear that the influence of this tiny dilution is minimal [see
Fig.~\ref{fig08}(a)] at this small $x$ value. This point was previously mentioned for
a specific $U$ ($= 12$) value in Fig.~\ref{fig07}.

Next, the question is: how does the ground state $U$-$t'$ phase diagram change as $x$
increases? To identify the different kinds of phases, first, we analyze the $U$-$t'$
phase diagram for $x = 0.1$ [see Fig.~\ref{fig08}(b)]. As we increase the percentage
of $U = 0$ sites, as compared to the $x = 0.02$ case, the phase diagram modifies
significantly. In particular, the system transitions from the G-type AF insulating
phase to the G-type AF metallic phase for $U > 6$ as $t'$ increases for $x = 0.1$ case.
In this dilution limit, intermediate $t'$ values promote percolating conduction between
$U = 0$ sites, converting the insulating G-type AF phase to metallic while maintaining
the underlying AF character. Furthermore, it is clear that the G-type AF metallic phase
stabilizes at a relatively smaller $t'$ value as $U$ increases; hence, the region over
which the G-type AF metallic phase is stabilized broadens with $U$ value. The G-type
AF metallic phase further transforms into C-type AF metallic phases as $t'$ increases,
either directly (for $U \ge 10$) or through a PM-M phase (from $8 < U < 10$).
The C-type AF phase is not stable for $U \le 8$ for $t' \le 1$. The phase diagram for
$x = 0.2$, shown in Fig.~\ref{fig08}(c), is qualitatively similar to that for $x = 0.1$.
Here, the G-type AF metallic region widens to some extent. Here, the G-type AF
metallic state also directly changes to the C-type AF metallic state with an
increase in $t'$, but only when $U \geq 16$. So, overall, the C-type AF region shrinks as
$x$ increases from $0.1$ to $0.2$.  This indicates that as dilution increases, the stability
of the C-type AF phase decreases. This C-type AF metallic phase completely disappears
from the phase diagram at $x = 0.4$ as shown in Fig.~\ref{fig08}(d), whereas the G-type
AF metallic phase remains stable in the intermediate $t'$ regime for $U > 6$. The
G-type AF metallic regime reduces slightly for $x = 0.4$ as compared to the $x = 0.2$,
then decreases considerably for $x = 0.6$ (see the inset) and vanishes completely
at higher $x$ ($\ge 0.8$) values.

\begin{figure}[t]
\centering
\includegraphics[width=0.48\textwidth]{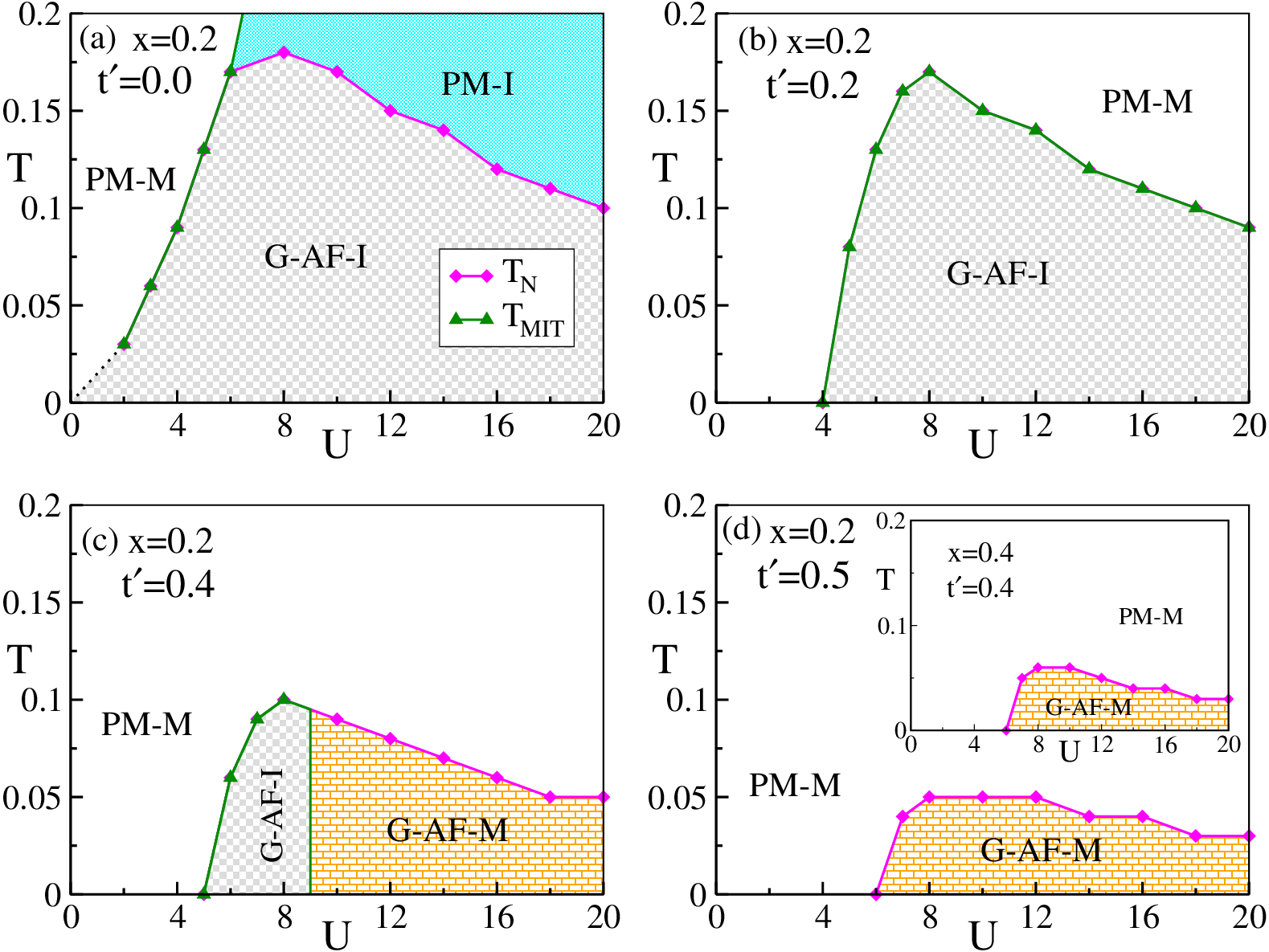}
\caption{The $U$-$T$ phase diagrams for $x = 0.2$ with various values of $t'$:
(a) $t' = 0$, (b) $t' = 0.2$, (c) $t' = 0.4$, and (d) $t' = 0.5$. Inset of (d) depicts 
the $U$-$T$ phase diagram with $x = 0.4$ and $t' = 0.4$.
(a) For $t' = 0$, the G-type AF insulating phase remains in the ground state
over the whole range of $U$ values. As the temperature increases, the AF phase
transitions to the PM-M phase for $U \le 6$ ($T_N = T_{MIT}$) and to the PM-I phase
for $U > 6$ ($T_N < T_{MIT}$). Similar to the non-diluted case~\cite{Mandal}, 
$T_N$ with $U$ is non-monotonic.
(b) As $t'$ increases to $0.2$, the long-range G-type AF order disappears for small
$U$ values ($U \le 4$), and the system remains in the PM-M state at low temperatures.
For $U > 4$, the G-type AF insulating phase is the ground state at low temperatures,
but the $T_N$ coincides with $T_{MIT}$ throughout the $U$ range. So, $T_N$ and
$T_{MIT}$ show the non-monotonicity with $U$. Note that when $t'$ increases, both
$T_N$ and $T_{MIT}$ decrease, but $T_{MIT}$ is significantly suppressed.
(c) The PM-M phase persists at low temperatures up to $U = 5$ for $t' = 0.4$. The
G-type AF insulating phase develops at low temperatures for only $5 < U < 9$. The
G-type AF metallic is interestingly the ground state
at low temperatures for $U > 9$. (d) The G-type AF phase that emerges for $U > 6$
turns out to be metallic for $t' = 0.5$. Therefore, the insulating region is
absent in this phase diagram. Inset: Even at $t' = 0.4$, the entire G-type AF
phase is shown to be metallic for $x = 0.4$. Legends are same in panels (a)$-$(d).
}
\label{fig09}
\end{figure}

To ascertain the variation $T_N$ of G-type AF (either insulating or metallic) phases
with $U$, we now focus on $x =  0.2$ and explore the $U$-$T$ phase diagrams for
four distinct $t'$ values. At $x = 0.2$, the $U$-$T$ phase diagram for $t' = 0$
[see Fig.~\ref{fig09}(a)] is very similar to the undiluted system~\cite{Mandal}. As
the temperature increases for $U \le 6$, the G-type AF insulator immediately transitions
into the PM-M phase. On the other hand, at finite temperatures, a PM-I phase separates from
the PM-M phase and the G-type AF insulator for $U > 6$. In contrast to $T_N$, which exhibits
non-monotonic behavior, $T_{MIT}$ manifests a monotonic increment, similar to the $x = 0$ case.
Increasing $t'$ from $0$ to $0.2$ significantly alters this scenario [see Fig.~\ref{fig09}(b)].
The finite temperature PM-I phase is suppressed, resulting in $T_N$ and $T_{MIT}$ coinciding
for all $U$ values given that G-type AF maintains the ground state (i.e., $U > 4$).
Thus, when $U > 4$, the high-temperature PM-M phase transitions to the G-type
AF insulating phase as the temperature decreases, and both $T_{MIT}$ and $T_N$ display
non-monotonic behavior with $U$. For $U\le 4$, the G-type AF order vanishes, due to the
addition of competing interactions, resulting in a low-temperature PM-M ground state.

The phase diagram becomes considerably more intriguing at $t' = 0.4$. Similar to the
$t' = 0.2$ scenario, the ground state stays PM metal at low $U$ ($\le 5$) values.
Interestingly, for $U > 5$, the system organized antiferromagnetically at low temperature;
yet, it exhibits metallicity through a large $U$ regime, that is, beyond $U = 9$
[see Fig.~\ref{fig09}(c)]. Thus, for $5 < U < 9$, $T_N$ coincides with $T_{MIT}$, where
the high-temperature PM-M phase transitions to the G-type AF insulating phase at low
temperatures, much like in the $t' = 0.2$ scenario. On the other hand, at low temperatures,
the PM-M phase switches into a G-type AF metallic phase for $U > 9$. The $T_N$ decreases
when $U$ increases beyond $U = 8$. As $t'$ grows to $0.5$, the entire AF
zone becomes metallic, as illustrated in Fig.~\ref{fig09}(d). But, the corresponding $T_N$
values decrease considerably for intermediate $U$ values. For $x = 0.4$ at $t' = 0.4$, a
very similar phase diagram is also obtained, where the entire G-type AF ordered region
has been found to be metallic [see inset Fig.~\ref{fig09}(d)].

\begin{figure}[t]
\centering
\includegraphics[width=0.48\textwidth]{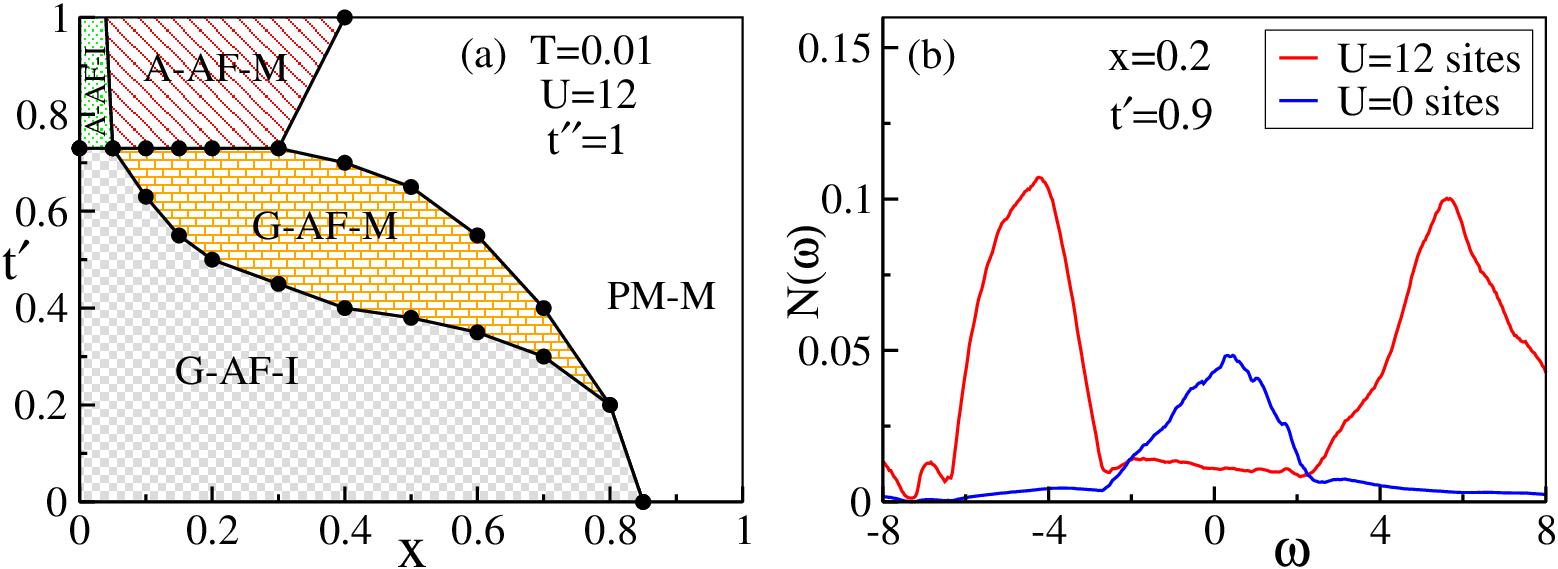}
\caption{Appearance of A-type AF metallic phase in the presence of the third-nearest-neighbor
hopping $t''$: (a) The $x$-$t'$ phase diagram with $t'' = 1$ depict that both the G-type
AF metallic and insulating get stabilized over a larger region in comparison to the
$t'' = 0$ scenario (see Fig.~\ref{fig07}). Notably, the upper left corner of the
phase diagram displays an A-type AF phase (large $t'$ and small to intermediate $x$ values).
The A-type AF state has been identified to be metallic, with the exception of the part that
remains close to $x = 0$. 
A-AF-I (A-AF-M) denotes the insulating (metallic) A-type AF phase in the phase diagram.
(b) The $U$-resolved DOS for $t' = 0.9$ and $x = 0.2$ exhibits
a noticeable peak near the Fermi level ($\omega = 0$), which is primarily contributed
by $U = 0$ sites. This suggests that the metallic behavior of A-type AF phase
in the phase diagram originates from the $U = 0$ sites.
}
\label{fig10}
\end{figure}

\section{Combined effect of long-range hopping ($t' \ne 0$ and $t'' \ne 0$)} \label{sec_ttt}

Our calculations thus far suggest that the addition of $t'$ hopping promotes the AF
metallic phases, composed of G- and C-type AF states in diluted systems (see Fig.~\ref{fig07}).
So, the A-type AF metallic phase is missing from our phase diagrams. Is it possible to
stabilize metallicity in A-type AF phases in diluted systems? It is worth noting that
the presence of $t''$ hopping stabilizes the A-type AF phase in undiluted systems~\cite{Mandal}.
We, therefore, examine the impact of third-nearest-neighbor hopping on the
magnetotransport properties of the diluted systems in order to identify potential
A-type AF metallic states. The ground state $x$-$t'$ phase diagram for $t'' = 1$
for $U = 12$ is explored in Fig.~\ref{fig10}(a) with the goal to contrast the results
with the $t'' = 0$ case. At $x \sim 0$ (i.e., for a nearly undiluted case), the
G-type AF insulating phase changes to an A-type AF insulating phase for large $t'$. 
As the dilution increases in this large $t'$ regime, the A-type AF 
insulating phase transforms into an A-type AF metallic phase. This A-type AF 
metallic phase remains stable up to $x \approx 0.3$. Therefore, a significant
difference from the $t'' = 0$ situation is the total suppression of the C-type
AF order and the emergence of the A-type AF phase in higher $t'$ values.

In addition, due to the cooperative nature of $t$ and $t''$, the G-type AF ordered
phase remains stable up to a comparatively larger value of $t'$ at $t'' = 1$ as
opposed to the $t'' = 0$ case. The G-type AF ordered region broadens and persists across
a wider range of $t'$ values at $t'' = 1$. This also expands the metallic region
associated with the G-type AF phase, which now appears at small to moderate values
of $t'$. As a result, it spans a larger portion of the phase diagram than that of the
$t'' = 0$ scenario. To analyze the role of $U = 0$ in the metallicity of the A-type
AF order we also plot $U$-resolved DOS for $x = 0.2$ (using $t' = 0.9$
and $t'' = 1$) at $T = 0.01$ in Fig.~\ref{fig10}(b). The $U$-resolved DOS clearly
shows that the $U = 0$ sites are the primary contributors to the finite density of
states at the Fermi level ($\omega = 0$). Thus, the $U = 0$ sites mainly assist 
the charge conduction, similar to the G- and C-type AF metallic phases.

\begin{figure}[t]
\centering
\includegraphics[width=0.48\textwidth]{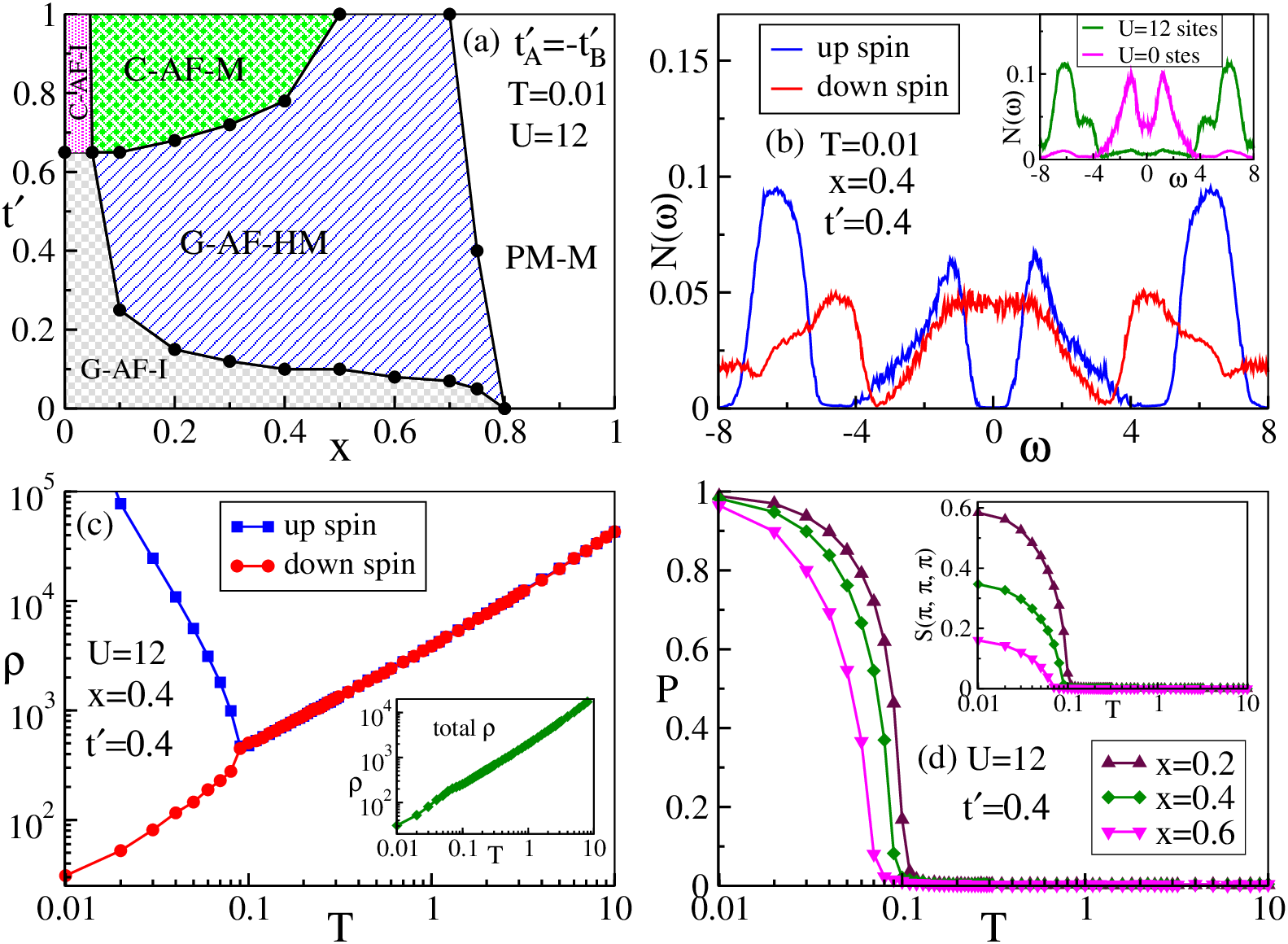}
\caption{Effect of sublattice-dependent second-nearest-neighbor hopping
($t' = t'_A = -t'_B$ where $A$ and $B$ are two sublattices):
(a) Interestingly, the $x$-$t'$ phase diagram depicts the emergence of a G-type
half-metallic phase. The G-type AF insulating phase is only present in
systems for small $x$ or small $t'$. A C-type AF metal phase, which
transitions to C-type AF insulating around $x = 0$, is observed at
large $t'$ and intermediate $x$ values. The ground state is determined to be
PM-M when the dilution is substantial ($x \ge 0.8$). 
G-AF-HM denotes the G-type antiferromagnetic half-metallic phase in the phase diagram.
(b) The spin-resolved DOS at $T = 0.01$ for $x = 0.4$ and $t' = 0.4$ for the
up-spin channel clearly reveals a gap at $\omega = 0$ (Fermi level), while the
down-spin channel exhibits a finite DOS. Inset: $U$-dependent DOS demonstrates
that $U = 0$ sites contribute to the down-spin channel in the main panel,
facilitating the conduction in the system.
(c) The temperature-dependent spin-resolved resistivities for $x = 0.4$ and
$t' = 0.4$ show a transition to a half-metallic state below the $T_N$.
The down-spin channel is conducting and the up-spin channel is insulating, which
is consistent with the DOS displayed in (c). In the inset total resistivity is
plotted with temperature.
(d) The temperature evolution of spin-polarization $P$ (main panel) and structure
factor $S(\pi,\pi,\pi)$ (inset) for $x = 0.2$, $0.4$, and $0.6$ for $t' = 0.4$ are
plotted. The onset of spin-polarization and the Neel temperature correspond well
for all $x$ values. The saturation value of $S(\pi,\pi,\pi)$ decreases at low
temperatures as the dilution increases, yet the system is found to be fully
spin-polarized ($P \sim 1$).
}
\label{fig11}
\end{figure}

\section{Sublattice-dependent Second-Nearest-Neighbor Hopping ($t' = t'_A = -t'_B \ne 0$): Engineering of half-metallic
antiferromagnets} \label{sec_tt12}

The spin polarization at the Fermi level, which depends on the difference in
the number of spin-up and spin-down electrons at the Fermi energy, plays a
crucial role in various physical phenomena and technological applications,
particularly in magnetic materials~\cite{Shim}. The higher the polarization,
the better it is for spintronic devices that rely on spin transport and spin injection
processes. The complete ($100\%$) spin polarization leads to a spin-dependent density
of states, with a gap for one spin orientation at the Fermi level (known as the
half-metallic gap) and a finite DOS for the opposite spin. This
peculiar electronic state---where a material can display metallic conductivity
for electrons of one particular spin orientation while acting as an insulator
or semiconductor for electrons of the opposite spin orientation---is the basis
for the concept of half-metallicity. Unfortunately, the AF
metallic systems that were identified by adjusting the $t'$ hopping parameters
in previous sections are not half-metallic. This is due to the fact that both
spin channels have a finite DOS at the Fermi level [see Fig.~\ref{fig06}].

The prospect of creating half-metallic antiferromagnets --- that is, systems
with full ($100\%$) spin polarization at the Fermi level --- will be discussed
in this section. We implement sublattice-dependent second-nearest-neighbor
hopping parameters, $t'_A$ and $t'_B$, to parameterize non-homogeneous hopping
in our calculations. We set $t' = t'_A = -t'_B$ and use third-nearest-neighbor
hopping $t'' = 0$ in our calculations. To compare to the homogeneous case, we
depict the ground state $x$-$t'$ phase diagram for $U = 12$ in Fig.~\ref{fig11}(a).
Notably, the phase diagram depicts that the metallic region exhibiting G-type
AF ordering is actually a half-metallic state. So, the system exhibits metallic
behavior in one spin polarization direction (e.g., spin-down) while insulating or
semiconducting in the other (e.g., spin-up).

To demonstrate the half-metallic character of the ground state, we explore
the spin-resolved DOS for $x = 0.4$ and $t' = 0.4$ in Fig.~\ref{fig11}(b) at low
temperature. It is evident that the up-spin channel exhibits a distinct gap at
the Fermi level ($\omega = 0$), while the down-spin channel displays a finite DOS. 
So, the imbalance in the hopping within sub-lattices triggers a mismatch in the
DOS at the Fermi level. In the inset of Fig.~\ref{fig11}(b), we also plot
the $U$-dependent DOS to show that the $U = 0$ sites primarily contribute
to the finite DOS at the Fermi level, which is analogous to the homogeneous
setting and provides the percolated conductive pathways in the system. Thus,
the conductive routes in this half-metallic case originate solely from
one spin channel of $U = 0$ sites.

Next, we plot the spin-resolved resistivity in Fig.~\ref{fig11}(c) to show the
spin-dependent transport properties of the half-metallic systems. Our 
calculations firmly show that only one spin channel---referred to as the down-spin
channel---participates in conduction because of the finite DOS around the Fermi
level in that specific spin channel. Thus, the resistivity of the up-spin channel
bifurcates with respect to the down-spin channel around $T_N$, even though the
total resistivity comprised of both spin channels of the systems exhibits metallic
nature at low temperatures [see the inset of Fig.~\ref{fig11}(c)]. The two spin
channels overlap above $T_N$, where the PM state is the ground state. The $T_N$ is
determined using the $S(\pi,\pi,\pi)$ vs. temperature plot in the inset of
Fig.~\ref{fig11}(d). We plot the temperature evolution of the spin-polarization
$P$ in Fig.~\ref{fig11}(d). The saturation value of spin-polarization at low
temperatures is $\sim 1$, demonstrating that at low temperatures, a $\sim 100\% $
spin-polarization is achieved. Please note that the temperature at which the upturn
is observed for $P$ corresponds to the $T_N$. Additionally, we use the same $t'$
to illustrate how spin polarization and $S(\pi,\pi,\pi)$ depend on temperature for
$x = 0.2$, $0.4$ and $0.6$ in Fig.~\ref{fig11}(d). Although the $T_N$ decreases with $x$,
the qualitative properties of half-metallic states remain consistent for all three
$x$ values.

Apart from the G-type AF phase, the polarization of the other magnetic phases
does not change when compared to the $t' = t'_A = t'_B$ setting [see Fig.~\ref{fig07}].
However, the size of the G-type insulating region on the phase diagram decreases
considerably. Comparatively, the G-type AF insulating phase is pushed to the
$t' \sim 0$ and $x \sim 0$ corner portions. Thus, it is apparent that the transition,
from the G-type AF insulating phase to a G-type AF half-metallic state, occurs at
comparatively small $t'$ ($t' = t'_A = -t'_B$) values and the half-metallic state
continues to be stable until a larger $t'$ value. Consequently, the area comprised
of the G-type AF half-metallic state turns out to be significantly broader. This
G-type AF half-metal transforms into a C-type AF metal without the spin polarization
signature when $t' \gtrsim 0.7$ and $x \lesssim 0.5$. Additionally, for large $t'$,
the C-type AF insulating phase appears in a restricted region close to $x \sim 0$.
Regardless of the $t'$ values, the long-range AF ordering vanishes at very large
dilutions ($x \ge 0.8$), and the high-temperature PM-M state remains the
ground state even at low temperatures.

\section{Conclusion}	\label{sec_con}

In conclusion, we addressed the central question of how introducing dilution into an
AF insulating system, in the presence of long-range magnetic interactions,
can induce metallicity. Our s-MC simulations on a diluted half-filled
one-band Hubbard model demonstrated that the site dilution, where the on-site repulsive Hubbard 
interaction strength is set to zero on a percentage of sites, leads to a weakening of the insulating
state. This weakening is attributed to percolative conduction among the diluted sites at low
temperatures, a remarkable phenomenon given the preservation of underlying long-range
AF ordering of the system. Furthermore, we illustrated how
sublattice-dependent hopping could potentially be utilized to the design of
half-metallic antiferromagnets, where the combined effect of site dilution and the competing
interactions plays a crucial role. Thus, our numerical results offer a comprehensive framework
for understanding the intricate interplay between site dilution and long-range interactions
in antiferromagnetic systems. These insights are essential, laying a fundamental basis that
will undoubtedly facilitate the rational design and development of novel AF
metals with tailored properties for advanced spintronic applications.

\section*{acknowledgment}
We acknowledge use of the Meghnad2019 computer cluster at SINP. 
We also acknowledge our discussions with Sandip Halder.

\appendix
\section{Effective Spin-Fermion Hamiltonian} \label{derivation_Heff}

The quartic interaction term in the model Hamiltonian [Eq.~\ref{hamiltonian}] is
decomposed into quadratic terms as mentioned in the main text. First, we express
the quartic interaction term as,
\begin{align}
H_I =   
n_{k,\uparrow} n_{k,\downarrow} = \frac{1}{4} n_k^2 - S_{kz}^2 = \frac{1}{4} n_k^2 - \left( \mathbf{S}_k.\hat{\Omega} \right)^2,
\end{align}
where $\mathbf{S}_k$ is the spin operator at site $k$ [defined as
$\mathbf{S}_k = \frac{1}{2} \sum_{\alpha\beta} c_{k,\alpha}^\dag \sigma_{\alpha\beta} c_{k,\beta}$, 
where $\hbar=1$ and $\sigma = (\sigma_x,\sigma_y,\sigma_z)$ represents the
Pauli matrices.] The term $\hat{\Omega}$ denotes an arbitrary unit vector. 
By exploiting the rotational invariance of $S_{kz}^2$, we use the relation
$\left( \mathbf{S}_k.\hat{\Omega} \right)^2 = S_{kx}^2 = S_{ky}^2 = S_{kz}^2$.
The partition function of the model Hamiltonian is expressed as
$Z = \mathrm{Tr}(\exp(-\beta H))$. Please note that $H = H_0 + H_I$ [see
Eq.~\ref{hamiltonian}]. The trace sums over all particle numbers and site
occupations. With $\beta = 1/T$ (setting $k_B = 1$), the interval $[0, \beta]$
is divided  into $M$ slices of width $\Delta \tau$, such that $\beta = M\Delta \tau$. 
Using the Suzuki-Trotter decomposition, for small $\Delta \tau$, we approximate
$\exp(-\beta (H_0 + H_I)) \approx \left( \exp(-\Delta \tau H_0) \exp(-\Delta \tau H_I) \right)^M$.
Applying the Hubbard-Stratonovich transformation, the interacting part of the partition
function $\exp[-\Delta \tau U \sum_k \{ \frac{1}{4} n_k^2 - ( \mathbf{S}_k.\hat{\Omega} )^2\}] $
for a given time slice ``$l$'' can be rewritten as
\begin{align}
&\int d \phi_k(l) \Delta_k(l) d^2\Omega_k(l) \exp \left[ - \Delta \tau \sum_k \left\{ \frac{\phi_k^2(l)}{U} + i\phi_k(l) n_k \right. \right. \nonumber \\ 
&\left. \left. + \frac{\Delta_k^2(l)}{U} -2\Delta_k(l) \hat{\Omega}_k(l) \cdot \mathbf{S}_k  \right\} \right].
\end{align}
The auxiliary fields $\phi_k(l)$ and $\Delta_k(l)$ are
introduced via the Hubbard-Stratonovich transformation. 
Defining $\mathbf{m}_k(l) = \Delta_k(l) \hat{\Omega}_k(l)$, 
we see that $\phi_k(l)$ couples to the local charge density, 
while $\mathbf{m}_k(l)$ couples to the local spin.  
Thus, the full partition function is proportional to:
\begin{align}
&\mathrm{Tr} \prod_{l=M}^1  \int d\phi_k(l) d^3\mathbf{m}_k(l) \exp \left[ - \Delta \tau \sum_k \left\{ \frac{\phi_k^2(l)}{U} + i\phi_k(l) n_k \right. \right. \nonumber \\ 
&\left. \left. + \frac{\mathbf{m}_k^2(l)}{U} -2\mathbf{m}_k(l) \cdot \mathbf{S}_k  \right\} \right].
\end{align}
The product order from $l=M$ to $l=1$ ensures time ordering. 
At this stage, the partition function remains exact, 
with the fields $\{ \phi_k(l), \mathbf{m}_k(l) \}$ fluctuating in both space and imaginary time.  
To proceed, we make the following approximations:  
(i) Neglecting the imaginary-time dependence of the auxiliary fields, keeping only their spatial variation.  
(ii) Using the saddle-point approximation $\phi_k(l) = iU \langle n_k \rangle /2$.  
With the redefinition $\mathbf{m}_k \to \frac{U}{2} \mathbf{m}_k$, 
these approximations lead to the effective spin-fermion Hamiltonian in Eq.~\ref{h_eff}, 
where fermions couple to classical auxiliary fields $\mathbf{m}_k$.

\section{Procedure to calculate observables} \label{obs}

We define the key observables to characterize magnetic and transport properties.
To determine the ground state magnetic ordering, we compute the quantum
correlations (magnetic structure factor) $S(\textbf{q})$ as follows:
\begin{align}
S(\textbf{q}) = \frac{1}{(L^3)^2} \sum_{kl} \left\langle \textbf{s}_k \cdot \textbf{s}_l \right\rangle \exp\left(i\textbf{q} \cdot (\textbf{r}_k - \textbf{r}_l) \right),
\end{align}
where $\mathbf{s}_k$ denotes quantum spin vector of site $k$.
The wave vector $\textbf{q}$ distinguishes different magnetic orders: 
$(\pi, \pi, \pi)$ for G-type AF; 
$(0, \pi, \pi), (\pi, 0, \pi), (\pi, \pi, 0)$ for C-type AF;
$(0, 0, \pi), (\pi, 0, 0), (0, \pi, 0)$ for A-type AF.
The summation runs over all sites with angular brackets denoting quantum and thermal averaging. 
The normalization factor $L^3$ represents the total number of sites.

The average local moment, representing the squared quantum magnetization, is
expressed as follows:
\begin{align}
M = \left\langle (n_{\uparrow} - n_{\downarrow})^2 \right\rangle = \left\langle n \right\rangle - 2 \left\langle n_{\uparrow} n_{\downarrow} \right\rangle .
\end{align}
Here $\left\langle n \right\rangle = \left\langle n_{\uparrow} + n_{\downarrow} \right\rangle $
denotes the average occupation of the electrons. In the extreme limit $U \to \infty$,
double occupancy becomes energetically forbidden, driving
$\left\langle n_{\uparrow} n_{\downarrow} \right\rangle$ to zero at any non-zero
temperature. If the system is half-filled, so that $\left\langle n \right\rangle = 1$,
the local moment approaches unity: $M \to 1$. Conversely, in the absence of
interactions ($U = 0$) or at very high temperatures ($T \to \infty$) for finite
$U$, electrons behave as uncorrelated particles. In this regime, double occupancy can
be approximated by the product
$\left\langle n_{\uparrow} \right\rangle \left\langle n_{\downarrow} \right\rangle$. 
At half-filling, this yields
$\left\langle n_{\uparrow} \right\rangle = \left\langle n_{\downarrow} \right\rangle = 0.5$,
leading to $M = 0.5$.
Therefore, under half-filling conditions, the local moment $M$ is bounded 
between 0.5 and 1, depending on the strength of the interactions and the
temperature.

For analyzing the transport properties, we compute the DOS and
the resistivity. The spin-resolved DOS is defined as follows:
\begin{align}
N_\sigma(\omega) = \sum_{i} \left\lvert \left\langle \sigma_i \vert \psi_i \right\rangle \right\rvert^2 \delta \left( \omega - \omega_i \right),
\end{align}
where $\omega_i$ and $\left\vert \psi_i \right\rangle$ are the eigenvalues and
eigenvectors of the effective Hamiltonian~\ref{h_eff}, respectively, and $\sigma_i$
denotes the spin of site $i$. 
To compute $N_\sigma(\omega)$, we employ a Lorentzian representation of the delta function, 
with a broadening of $\sim BW/(2L^3)$, where $BW$ represents the non-interacting bandwidth 
and $L^3$ is the total number of lattice sites. The total DOS is computed as follows: 
\begin{align}
N(\omega) &= \sum_\sigma N_\sigma(\omega) = \sum_{i, \sigma} \left\lvert \left\langle \sigma_i \vert \psi_i \right\rangle \right\rvert^2 \delta \left( \omega - \omega_i \right) \nonumber \\ 
&= \sum_{i=1}^{2L^3} \delta(\omega - \omega_i),
\end{align}
where $\sum_\sigma \left\vert \sigma_i \right\rangle = \left\vert \psi_i \right\rangle$.

The spin polarization is defined in terms of the spin-resolved DOS at the Fermi level ($\omega = 0$) as
\begin{align}
P = \frac{N_\uparrow(0) - N_\downarrow(0)}{N_\uparrow(0) + N_\downarrow(0)},
\end{align}
where $N_\uparrow(0)$ and $N_\downarrow(0)$ denote the DOS at $\omega = 0$ for up and down spins, respectively.

The resistivity is calculated by inverting the $dc$ limit of the optical conductivity, 
obtained using the Kubo-Greenwood formula~\cite{Mahan, Kumar2}.
We compute the optical conductivity as follows:
\begin{align}
\sigma(\omega) = \frac{\pi e^2}{N \hbar a_0} \sum_{\alpha, \beta, \sigma} \left( n_\alpha -n_\beta \right) \frac{\left\lvert f_{\alpha \beta}^\sigma \right\rvert}{\omega_\beta - \omega_\alpha} \delta \left( \omega - \left( \omega_\beta - \omega_\alpha \right) \right),
\end{align}
where $f_{\alpha \beta}^\sigma=\left\langle \psi_\alpha \left\vert \hat{j}_{z,\sigma} \right\vert \psi_\beta \right\rangle$ 
are the matrix elements of the spin-dependent current operator 
$\hat{j}_{z,\sigma}=ia_0 \sum_{i} [t ( c_{i,\sigma}^\dag c_{i+a_0\hat{z},\sigma} -h.c.)+t'( c_{i,\sigma}^\dag c_{i+a_0\hat{z}+a_{0}\hat{x},\sigma} -h.c.)+t'( c_{i,\sigma}^\dag c_{i+a_0\hat{z}+a_{0}\hat{y},\sigma} -h.c.)+t''( c_{i,\sigma}^\dag c_{i+a_0\hat{x}+a_0\hat{y}+a_0\hat{z},\sigma} -h.c.)]$.
Here, $\omega_\alpha$ and $\left\vert \psi_\alpha \right\rangle$ 
are the eigenvalues and associated eigenvectors of the effective Hamiltonian~\ref{h_eff}, and 
$n_\alpha = f(\mu - \omega_\alpha)$ are Fermi factors.
The average $dc$ conductivity is computed by integrating over a small frequency window
($\Delta\omega$): $\sigma_{dc} = (\Delta\omega)^{-1} \int_0^{\Delta\omega} \sigma(\omega) \, d\omega$. 
For the spin-resolved conductivity, $\sigma_{dc,\sigma}$, 
spin-resolved states and operators are used in constructing 
$f_{\alpha \beta}^\sigma=\left\langle \psi_\alpha \left\vert \hat{j}_{z,\sigma} \right\vert \psi_\beta \right\rangle$ 
in the conductivity expression.
Finally, resistivity is obtained as the inverse of the average $dc$ conductivity.



\end{document}